\documentclass{nature_sub}

\usepackage{graphicx}
\usepackage{amsmath}
\usepackage{amssymb}
\usepackage{color}
\usepackage{hyperref}
\usepackage{verbatim}

\newcommand{\apj}{Astrophys. J.}
\newcommand{\apjs}{Astrophys. J. Suppl.}
\newcommand{\apjl}{Astrophys. J. Lett.}
\newcommand{\aap}{Astron. Astrophys.}
\newcommand{\aj}{Astron. J.}
\newcommand{\mnras}{Mon. Not. R. Astron. Soc.}

\newcommand{\prd}{Phys. Rev. D}

\newcommand{\prl}{{Phys. Rev. Lett.}}

\newcommand{\nat}{{Nature}}

\newcommand{\physrep}{Phys. Rep.}

\newcommand{\pthphys}{Prog. Theor. Phys.}

\newcommand       \atan       {{\rm atan}}

\newcommand       \acosh       {{\rm acosh}}

\usepackage{multibib}

\newcites{main,meth}%
         {,%
          }
\bibliographystylemain{naturemag}
\bibliographystylemeth{naturemag}

\title{A Statistical Solution to the Chaotic, Non-Hierarchical Three-Body Problem}

\author{Nicholas C. Stone$^{1,2,3}$, Nathan W.C. Leigh$^{4,5}$}

\begin{document}

\maketitle

\begin{affiliations}
 \item Columbia Astrophysics Laboratory, Columbia University, New York, NY 10027, USA
 \item Racah Institute of Physics, The Hebrew University, Jerusalem, 91904, Israel
 \item Department of Astronomy, University of Maryland, College Park, MD, 20742, USA
 \item Department of Astrophysics, American Museum of Natural History, Central Park West and 79th Street, New York, NY, 10024, USA
 \item Departamento de Astronom\`ia, Facultad de Ciencias F\`isicas y Matem\`aticas, Universidad de Concepci\`on, Chile
\end{affiliations}

\begin{abstract}
The three-body problem is arguably the oldest open question in astrophysics, and has resisted a general analytic solution for centuries.  Various implementations of perturbation theory provide solutions in portions of parameter space, but only where hierarchies of masses or separations exist.  Numerical integrations\cite{Agekyan&Anosova67} show that bound, non-hierarchical triples of Newtonian point particles will almost\cite{Suvakov&Dmitrasinovic13} always disintegrate into a single escaping star and a stable, bound binary\cite{Standish72, HutBahcall83}, but the chaotic nature of the three-body problem\cite{Poincare92} prevents the derivation of tractable\cite{Sundman12} analytic formulae deterministically mapping initial conditions to final outcomes.  However, chaos also motivates the assumption of ergodicity\cite{Fermi50, Monaghan76a, Valtonen+05}, suggesting that the distribution of outcomes is uniform across the accessible phase volume.  Here, we use the ergodic hypothesis to derive a complete statistical solution to the non-hierarchical three-body problem, one which provides closed-form distributions of outcomes (e.g. binary orbital elements) given the conserved integrals of motion.  We compare our outcome distributions to large ensembles of numerical three-body integrations, and find good agreement, so long as we restrict ourselves to ``resonant'' encounters\cite{Heggie75} (the $\sim 50\%$ of scatterings that undergo chaotic evolution).  In analyzing our scattering experiments, we identify ``scrambles'' (periods in time where no pairwise binaries exist) as the key dynamical state that ergodicizes a non-hierarchical triple.  The generally super-thermal distributions of survivor binary eccentricity that we predict have notable applications to many astrophysical scenarios.  For example, non-hierarchical triples produced dynamically in globular clusters are a primary formation channel for black hole mergers\cite{PortegiesZwart&McMillan00, Rodriguez+16, Hong+18}, but the rates and properties\cite{Samsing+14, Rodriguez+18} of the resulting gravitational waves depend on the distribution of post-disintegration eccentricities.
\end{abstract}

The three-body problem is a prototypical example of deterministic chaos\cite{Poincare92}, in that tiny perturbations in initial conditions (or errors in numerical integration) lead to exponentially divergent outcomes\cite{PortegiesZwart&Boekholt18}.  
Chaotic systems often forget their initial conditions (aside from integrals of motion), though this is by no means guaranteed, and indeed, the topology of the chaotic three-body problem does contain islands of regularity\cite{Hut83, SamsingIlan17}.  Nonetheless, to a first approximation, it is reasonable to estimate the probability of different outcomes by invoking the ergodic hypothesis\cite{Bohr36, Fermi50}, and to assume that non-hierarchical triples will uniformly explore the phase space volume accessible to them\cite{Monaghan76a}.  In this way, we may turn the chaotic nature of the three-body problem\cite{Poincare92, PortegiesZwart&Boekholt18} - which has, so far, frustrated general, deterministic, analytic mappings from one set of initial conditions to one set of outcomes - into a tool that simplifies the mapping from {\it distributions} of initial conditions to {\it distributions} of outcomes.  

Consider the generic outcome of the non-hierarchical Newtonian three-body problem: a single escaper star, with mass $m_{\rm s}$, departs from a surviving binary with mass $m_{\rm B} = m_{\rm a} + m_{\rm b}$, where $m_{\rm a}$ and $m_{\rm b}$ are the component masses.  The binary components are separated by a distance $\vec{r}$ and have relative momentum $\vec{p}$, while the escaper is separated from the binary center of mass by $\vec{r}_{\rm s}$ and is moving with relative momentum $\vec{p}_{\rm s}$.  The total energy and angular momentum of the system, inherited from the initial conditions and preserved through a period of chaotic three-body interactions, are $E_0$ and $\vec{L}_0$, respectively.  For convenience, we define additional masses $M=m_{\rm s}+m_{\rm B}$, $m = m_{\rm B}m_{\rm s}/M$, and $\mathcal{M}=m_{\rm a}m_{\rm b}/m_{\rm B}$.  The total accessible phase volume for this system is that of an 8-dimensional hypersurface\cite{Monaghan76a}:
\begin{equation}
\sigma = \idotsint \delta(E_{\rm B}+E_{\rm s}-E_0)\delta(\vec{L}_{\rm B}+\vec{L}_{\rm s}-\vec{L}_{0}){\rm d}\vec{r}{\rm d}\vec{p}{\rm d}\vec{r}_{\rm s}{\rm d}\vec{p}_{\rm s}, \label{eq:sigmaCart}
\end{equation}
shaped by the requirements of energy and angular momentum conservation for both the elliptic orbit of the surviving binary ($E_{\rm B}$, $\vec{L}_{\rm B}$) and the hyperbolic orbit between the binary and the escaper ($E_{\rm s}$, $\vec{L}_{\rm s}$).  Given a microcanonical ensemble of non-hierarchical triples with different initial conditions but identical integrals of motion and mass combinations, the outcome states (after breakup) will - assuming ergodicity - uniformly populate the phase volume accessible at the moment of disintegration.  This ensemble is microcanonical in the sense that each three-body system is isolated from external sources of heat, but is unusual in its low particle number\cite{Fermi50}.

We evaluate this integral at the moment of disintegration, which we idealize as occuring anywhere inside a ``strong interaction region'' of radius $R(E_{\rm B}, L_{\rm B}, C_{\rm B})$, where $C_{\rm B} = \hat{L}_{\rm B}\cdot \hat{L}_0$.  Canonical transformations to elliptic/hyperbolic Delaunay elements facilitate the integration (see Supplementary Information) and yield a phase volume of
\begin{align}
\sigma&=\frac{2\pi^4G^2 M^{5/2} m_{\rm B} }{(m_{\rm a} m_{\rm b} m_{\rm s})^{3/2}} \iiint \frac{L_{\rm B}{\rm d}E_{\rm B}{\rm d}L_{\rm B}{\rm d}C_{\rm B}}{L_{\rm s}(-E_{\rm B})^{3/2}(E_0 - E_{\rm B})^{3/2}} \notag \\
& \times \Bigg(\sqrt{\frac{2M(E_0-E_{\rm B})}{G^2m_{\rm s}^3 m_{\rm B}^3} } \sqrt{ 2m(E_0 - E_{\rm B}) R^2 + 2GMm^2R - L_{\rm s}^2} \notag \\
& -\acosh\left( \frac{1+2(E_0 - E_{\rm B})R/(Gm_{\rm s}m_{\rm B})}{\sqrt{1+2M(E_0 - E_{\rm B})L_{\rm s}^2/(G^2m_{\rm s}^3m_{\rm B}^3)}} \right) \Bigg) \label{eq:solution2}.
\end{align}
For brevity, we have re-inserted the angular momentum of the escaping star, $L_{\rm s}^2(L_{\rm B}, C_{\rm B}) \equiv L_{\rm B}^2(1-C_{\rm B}^2) + (L_{\rm B}C_{\rm B} - L_0)^2$.
While $\sigma$ is a phase volume, the integrand of Eq. \ref{eq:solution2} is a trivariate outcome distribution representing the differential probability of finding a disintegrating metastable triple in a volume ${\rm d}E_{\rm B}{\rm d}L_{\rm B}{\rm d}C_{\rm B}$: the microcanonical ensemble for survivor binaries produced in the non-hierarchical three-body problem (other, angular, binary orbital elements are distributed uniformly). 
Specification of total energy $E_0$ and total angular momentum $\vec{L}_0$ suffices, therefore, to describe the {\it distribution} of outcomes in non-hierarchical triple systems, even if this information alone cannot deterministically specify how one {\it individual} outcome follows from one set of initial conditions.  Conservation of $E_0$ and $\vec{L}_0$ means that the trivariate outcome distribution in Eq. \ref{eq:solution2} can be mapped one-to-one to the distribution of escaper properties.  Eq. \ref{eq:solution2} makes fewer simplifying assumptions than did past ergodic analyses of the general three-body problem\cite{Monaghan76a, Monaghan76b, Nash&Monaghan78, Valtonen+05}, and its outcome distributions are qualitatively different.

We marginalize over $L_{\rm B}$ and $C_{\rm B}$ to compute the distribution of outcome energies, ${\rm d}\sigma / {\rm d}E_{\rm B}$.  In an $L_0 = 0$ ensemble, this is ${\rm d}\sigma / {\rm d}E_{\rm B} \propto |E_{\rm B}|^{-7/2}$, extending to $|E_{\rm B}| \to \infty$.  Conversely, when $L_0$ is large, the ergodic energy distribution is slightly steeper, going roughly as ${\rm d}\sigma / {\rm d}E_{\rm B} \propto |E_{\rm B}|^{-4}$, but only out to a maximum energy $|E_{\rm max}| \propto L_0^{-2}$; larger outcome energies are prohibited by angular momentum conservation.  The energy distribution we calculate differs from past estimates determined assuming detailed balance\cite{Heggie75}, demonstrating that a population of binaries engaging in ergodic three-body interactions with a thermal bath of single stars {\it cannot achieve detailed balance}, so long as their outcomes are ergodically distributed.

We likewise integrate to find the marginal outcome distributions in angular momentum (which we represent in terms of binary eccentricity $e_{\rm B}$, as ${\rm d}\sigma/{\rm d}e_{\rm B}$) and inclination (${\rm d}\sigma / {\rm d}C_{\rm B}$).  In contrast to the usual (though not universal\cite{Geller+19}) expectation of a thermal eccentricity distribution, ${\rm d}\sigma/{\rm d}e_{\rm B} = 2e_{\rm B}$, we find a mildly super-thermal eccentricity distribution for large $L_0$: ${\rm d}\sigma/{\rm d}e_{\rm B} = \frac{6}{5}e_{\rm B}(1+e_{\rm B})$.  This radial orbit bias is a geometric effect arising from the larger average interaction cross-section of a highly eccentric binary, the apocenter of which is twice as large as that of a circular binary of equal energy.  
In the low-$L_0$ limit, the ergodic distribution of survivor eccentricities is {\it highly super-thermal}, with ${\rm d}\sigma/{\rm d}e_{\rm B} \propto e_{\rm B}(1+e_{\rm B})/\sqrt{1-e_{\rm B}^2}$ when $L_0 = 0$.  There is a strong bias towards producing nearly radial binaries, as a consequence of {\it angular momentum starvation}:  while a low-$L_0$ ensemble of non-hierarchical triples may produce a quasi-circular survivor binary, doing so requires substantial fine-tuning of the angle and velocity of the escaper, and is therefore disfavored.  Similar phase volume considerations explain the strong bias towards prograde ($0<C_{\rm B}\le 1$) orbits Eq. \ref{eq:solution2} predicts when marginalized into ${\rm d}\sigma/{\rm d}C_{\rm B}$.  More detailed explorations of the ergodic ${\rm d}\sigma / {\rm d}E_{\rm B}$, ${\rm d}\sigma / {\rm d}e_{\rm B}$, and ${\rm d}\sigma / {\rm d}C_{\rm B}$ distributions are shown in Extended Data Figs. \ref{fig:marginalEnergy}, \ref{fig:marginalEccentricity}, and \ref{fig:marginalInclination}, respectively, as well as in the Supplementary Information.

Our outcome distribution, ${\rm d}\sigma/{\rm d}E_{\rm B}{\rm d}L_{\rm B}{\rm d}C_{\rm B}$, was derived with several assumptions, most notably: (i) the ergodic hypothesis; (ii) instantaneous disintegration; (iii) a specific parametrization of the ``strong interaction region'' defining the limits of integration.  It should therefore be tested against ensembles of numerical scattering experiments.  We have explored the ergodicity of non-hierarchical triples in the equal-mass limit, by using the \texttt{FEWBODY} numerical scattering code to run three ensembles of different binary-single scattering experiments (see Extended Data Table 1).  Each ensemble has roughly $N\approx 10^5$ runs with constant $E_0$ and $L_0$, but otherwise random initial conditions.  However, many of our scattering experiments do not form resonant three-body systems, but instead resolve abruptly in a prompt exchange, where it is unlikely that the ergodic hypothesis can be applied.  Metastable three-body systems generally exhibit intermittent chaos\cite{Pomeau&Manneville80}.  Long periods of quasi-regular evolution occur during the non-terminal ejection of a single star, but these are then interrupted by brief periods of intensely chaotic evolution when that star returns to pericenter\cite{Heggie75, HutBahcall83}.  We hypothesize that the degree of ergodicity in a subset of scattering experiments can be inferred from the number of ``scrambles,'' $N_{\rm scram}$: periods of time when no pairwise binary exists.  

\begin{figure}
\includegraphics[width=173mm]{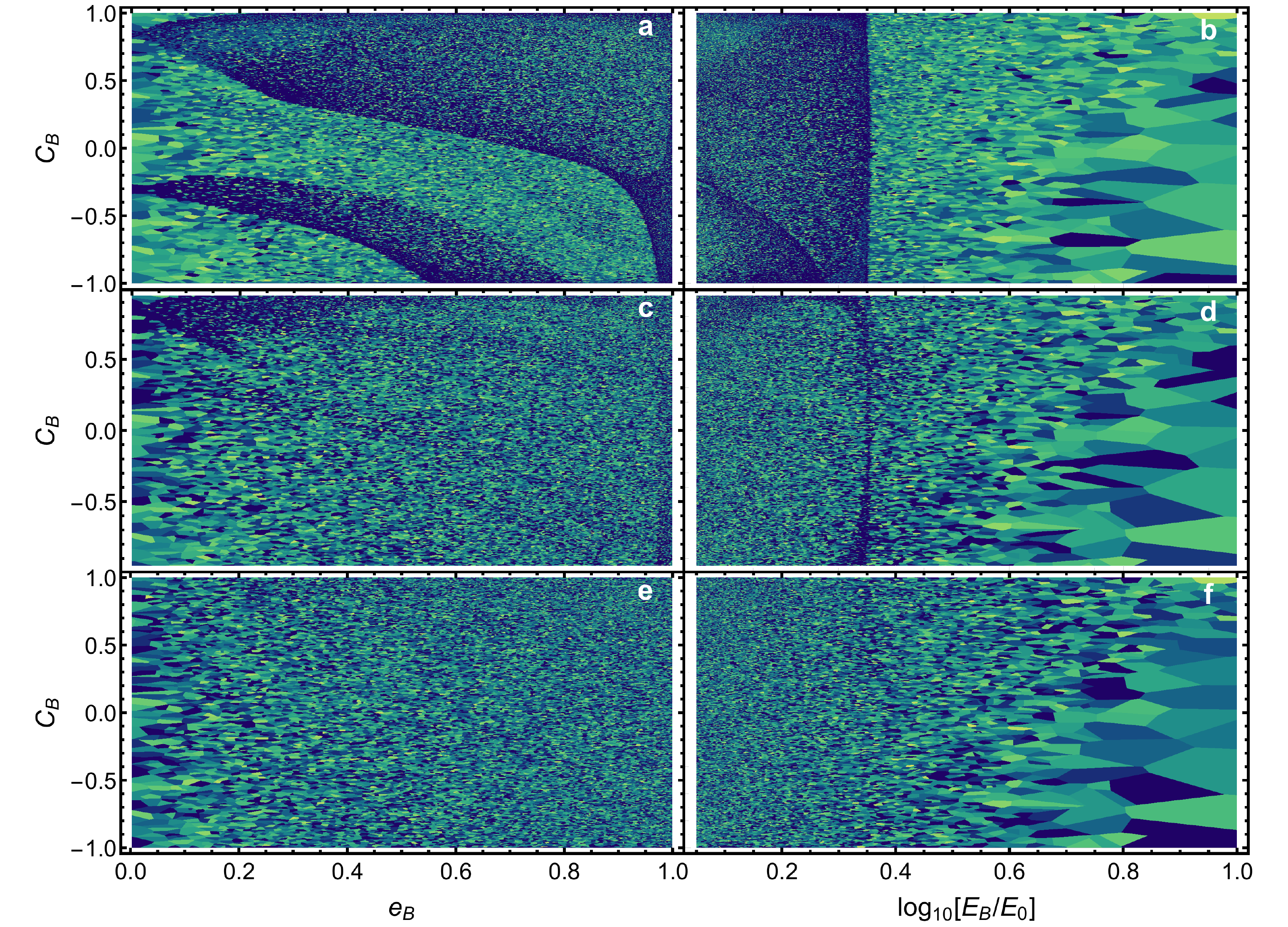}
\caption{{\bf Topological maps of three-body scattering outcomes for Run A}.  The total number of scrambles is color-coded (smallest values of $N_{\rm scram}$ as dark blue, larger $N_{\rm scram}$ in green and yellow) with a logarithmic scaling, as a function of survivor binary eccentricity $e_{\rm B}$ (panels {\bf a, c, e}), energy $E_{\rm B}$ (panels {\bf b, d, f}) and cosine-inclination $C_{\rm B}$.  Different panels show $N_{\rm scram} \ge 0$ ({\bf a, b}), $N_{\rm scram} \ge 1$ ({\bf c, d}), and $N_{\rm scram} \ge 2$ ({\bf e, f}).  Clouds of regularity obscure the underlying chaotic sea in the top two panels, but have dissipated in the bottom panel, indicating that scrambles are the key dynamical mechanism responsible for ``ergodicizing'' the comparable-mass three-body problem.}
\label{fig:ergodicDevelopment}
\end{figure}

\begin{figure}
\includegraphics[width=165mm]{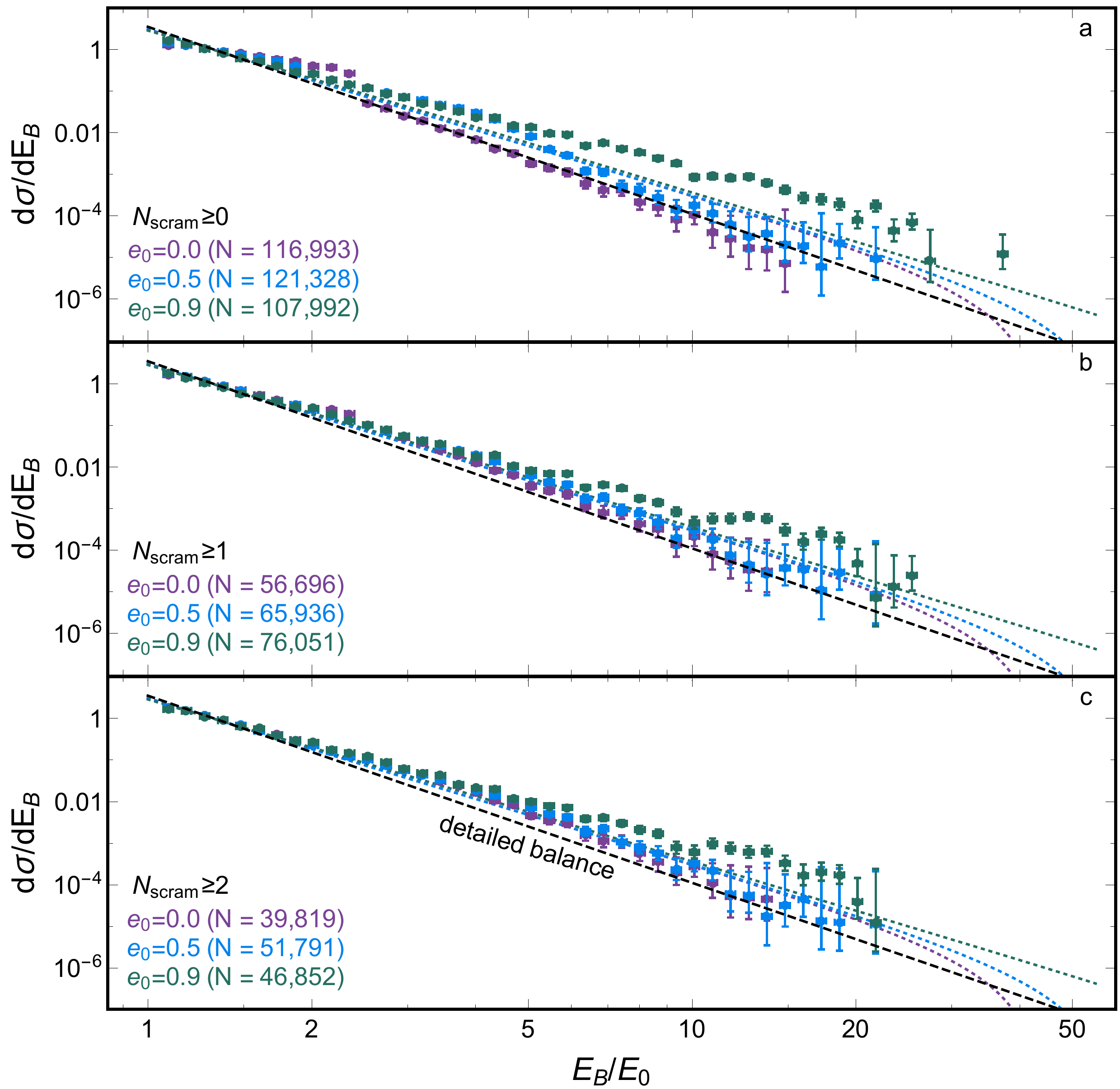}
\vspace{-45pt}
\caption{{\bf The marginal distribution of binary energy}, ${\rm d}\sigma/{\rm d}E_{\rm B}$, plotted against dimensionless energy $E_{\rm B}/E_0$.  The dotted lines are ergodic outcome distributions for high (purple), medium (blue), and low (green) angular momentum ensembles.  The data points are binned outcomes from numerical binary-single scattering ensembles ($N\approx10^5$).  {\bf a}: the full set of results from our numerical scattering experiments.  {\bf b}: the subset of results where the number of scrambles, $N_{\rm scram} \ge 1$.  {\bf c}: the subset of results where $N_{\rm scram}\ge2$.  Detailed balance (black dashed line) is never achieved.}
\label{fig:marginalEnergyNumeric}
\end{figure}

\begin{figure}
\includegraphics[width=175mm]{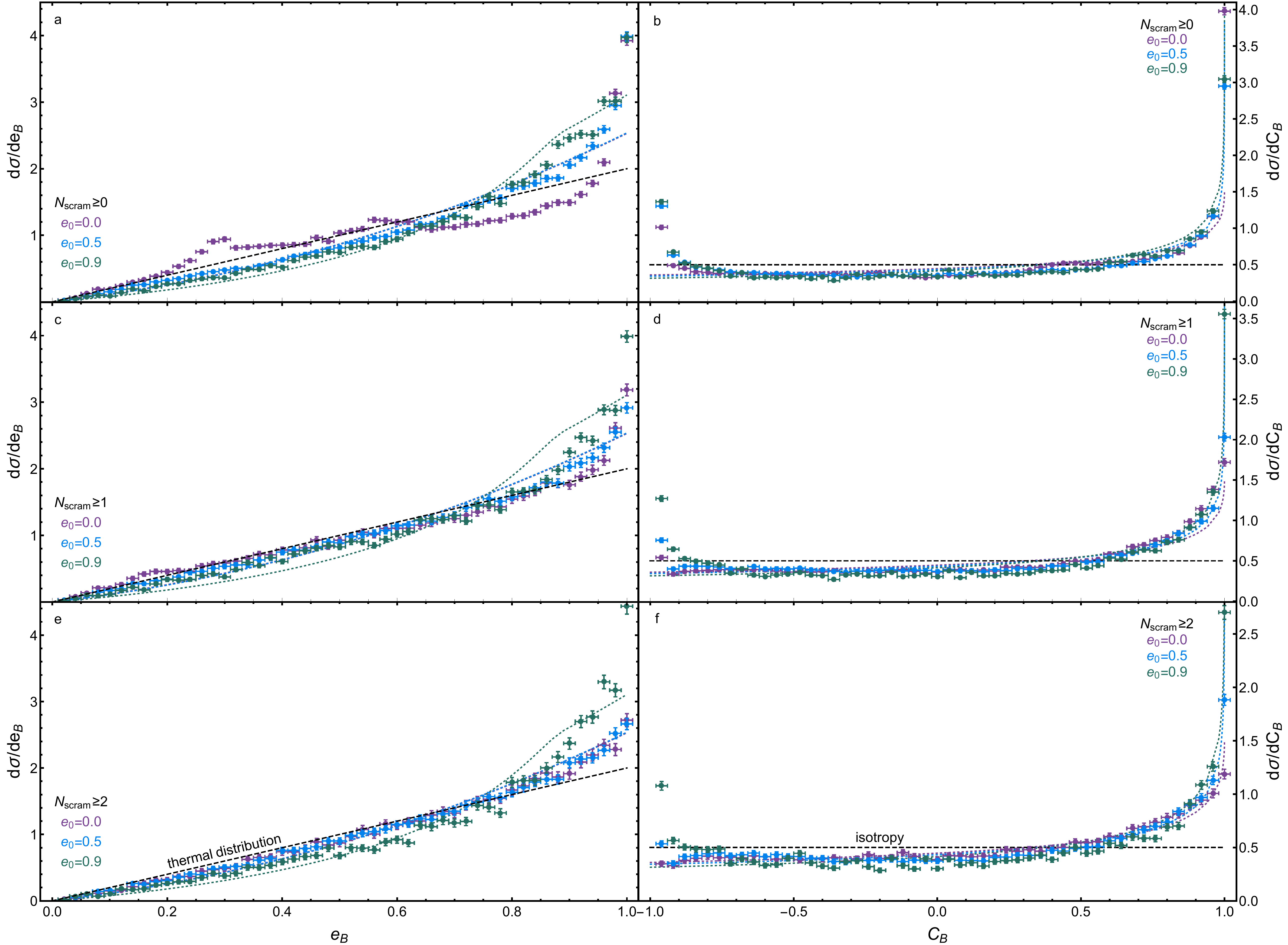}
\caption{{\bf The marginal distributions of binary eccentricity and orientation.}  Panels {\bf a, c, e}: ${\rm d}\sigma/{\rm d}e_{\rm B}$ plotted against eccentricity $e_{\rm B}$.  Panels {\bf b, d, f}: ${\rm d}\sigma/{\rm d}C_{\rm B}$ plotted against the cosine of the binary inclination, $C_{\rm B}$.  Line styles represent ergodic outcome distributions with the same ensemble angular momenta as in Fig. \ref{fig:marginalEnergyNumeric}.  The data points are binned outcomes from the same numerical scattering ensembles as in Fig. \ref{fig:marginalEnergyNumeric}, with each row corresponding to the same cuts on $N_{\rm scram}$.  Eccentricity outcome distributions are notably super-thermal (the thermal distribution ${\rm d}\sigma / {\rm d}e_{\rm B} = 2e$ is shown as a black dashed line).  Inclination distributions exhibit anisotropic bias towards prograde binaries aligned with $\vec{L}_0$ (the isotropic distribution is shown with a black dashed line).}
\label{fig:marginalMultiNumeric}
\end{figure}

We illustrate the development of ergodicity in Fig. \ref{fig:ergodicDevelopment}, which shows topological maps in outcome space.  While the full scattering ensemble has clear geometrical features indicative of prompt exchanges, these ``clouds of regularity'' mostly (entirely) disappear if one considers the $\approx 50\%$ of integrations with $N_{\rm scram}\geq 1$ ($N_{\rm scram}\geq 2$).  With this qualitative argument in mind, we now use Figs. \ref{fig:marginalEnergyNumeric} and \ref{fig:marginalMultiNumeric} to quantitatively compare the binned results of our scattering experiments to the marginal distributions predicted by the ergodic hypothesis.  Horizontal error bars show bin sizes, and vertical error bars indicate $95\%$ Poissonian confidence intervals.  All three of the marginal distributions we examine (${\rm d}\sigma/{\rm d}E_{\rm B}$, ${\rm d}\sigma/{\rm d}e_{\rm B}$, and ${\rm d}\sigma/{\rm d}C_{\rm B}$) exhibit reasonable (and sometimes very close) agreement between the ergodic theory of Eq. \ref{eq:solution2} and our numerical scattering experiments, provided we examine resonant encounters ($N_{\rm scram} \geq 2$).  The marginal distributions for large-${L}_0$ ensembles are in very good agreement with the numerical experiments.  The agreement is slightly worse for our low-${L}_0$ ensemble.

The agreement between ergodic theory and experiment is never exact, even in $N_{\rm scram} \ge 2$ subsamples, and in most cases we see data that matches analytic predictions to leading order, but also exhibits some level of higher-order structure.  The nature of these superimposed, second-order structures is not altogether clear, as two explanations seem plausible.  First, these could represent islands of regularity in the initial conditions we have explored: regions of parameter space that do not fully forget their initial conditions despite undergoing multiple scrambles.  Second, these could represent a failure in the idealized escape criteria, $R(E_{\rm B}, L_{\rm B})$, that we employ.  We have only considered very simple definitions of the strong interaction region, the true shape of which is likely connected to the triple stability boundary\cite{Mardling&Aarseth01}.  We defer an investigation of these two hypotheses to future work.

Non-hierarchical triples are common, if short-lived, in the astrophysical Universe\cite{Leigh&Geller13}.  They are responsible for many interesting phenomena.  For example, binary-single scattering events in dense star clusters produce blue stragglers\cite{Leonard&Fahlman91, Leigh+11}, cataclysmic variables\cite{Ivanova+06}, X-ray binaries\cite{Pooley&Hut06, Ivanova+08}, and even binary stellar-mass black holes\cite{PortegiesZwart&McMillan00}.  The lattermost of these scenarios may be responsible for most of the black hole mergers seen by the LIGO experiment\cite{Rodriguez+16, Hong+18}.  Dynamical formation of these systems in a binary-single scattering is favored when the surviving binary is drawn from the high-$e_{\rm B}$ tail of outcomes.  It is therefore notable that (i) we find generic superthermality in the outcomes of comparable-mass scatterings (both from ergodic theory and numerical experiments), and (ii) that our formalism has identified the type of binary-single encounters that are predisposed to produce exotic binaries: {\it low-$L_0$ scatterings}.  In the future, it may be possible to apply our formalism to estimate the properties of temporary binaries formed during long, but non-terminal, ejections of the single star.  High eccentricity binaries formed as ``intermediate states'' of a three-body resonance may merge during the ejection due to short-range dissipative forces, leading to, e.g., uniquely eccentric gravitational wave signals\cite{Samsing+14}.

\begin{addendum}
 \item[Acknowledgments] We gratefully acknowledge useful discussions with Douglas Heggie, Piet Hut, Re'em Sari, and Simon Portegies-Zwart, as well as constructive feedback from two anonymous referees.  N.C.S. received financial support from NASA, through both Einstein Postdoctoral Fellowship Award Number PF5-160145 and the NASA Astrophysics Theory Research Program (Grant NNX17AK43G; PI B. Metzger); he also thanks the Aspen Center for Physics for its hospitality during early stages of this work.  N.W.C.L. gratefully acknowledges the generous support of a Fondecyt Iniciacion grant, \#11180005. Both authors thank the Chinese Academy of Sciences for hosting us as we completed our efforts.  We extend special thanks to Mauri Valtonen and Hanno Karttunen, whose superb book on the three-body problem motivated much of this work.
 \item[Author Contributions] N.C.S. led the analytic work, which N.W.C.L. contributed significantly to.  The \texttt{FEWBODY} simulations were performed by N.W.C.L.  The comparison between the simulations and the analytic theory was jointly split between the two authors.
 \item[Author Information] Reprints and permissions information is available at www.nature.com/reprints. The authors declare no competing financial interests. Correspondence and requests for materials should be addressed to N.C.S.~(nicholas.stone@mail.huji.ac.il).
\end{addendum}

\newpage

\begin{figure}
\centering
\includegraphics[width=89mm]{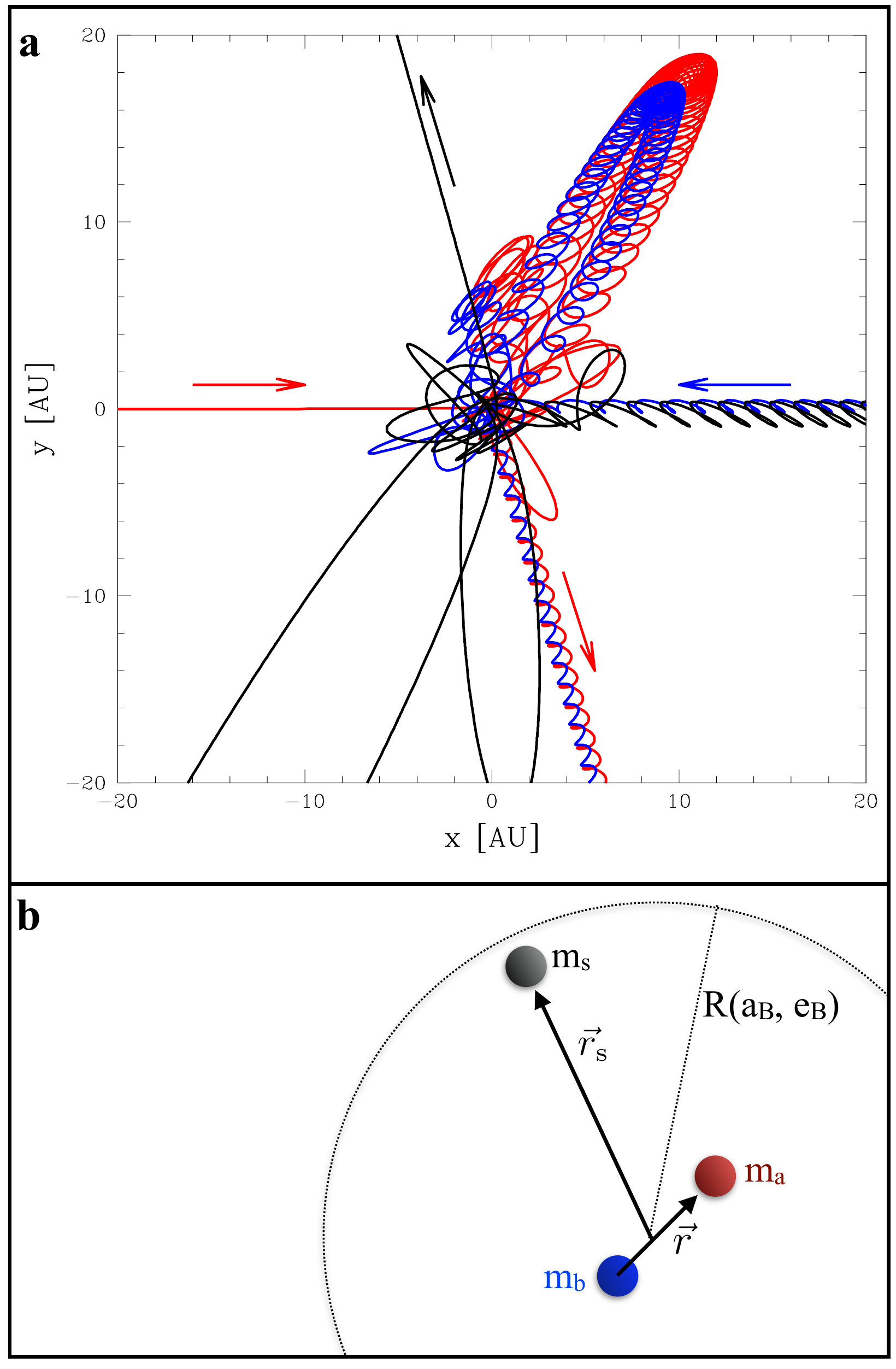}
\caption{{\bf Non-hierarchical three-body scatterings.}  {\bf a}: the two dimensional projection of an equal-mass resonant scattering encounter, where an interloper star (red) encounters a binary (blue and black).  The resonant interaction unfolds over several dynamical times before the system disintegrates in a partner swap.  {\bf b}: a schematic illustration of the metastable triple at the moment of disintegration.}
\label{fig:cartoon}
\end{figure}

\clearpage

\begin{methods}

\section{Chaotic Escape in the Three-Body Problem}
\label{sec:derivation}

Unlike the two-body problem, which admits closed-form solutions, the general three-body problem is substantially more complex.  Analytic treatments exist in {\it hierarchical} regimes, where the masses or separations of the three bodies differ greatly, but, aside from certain measure-zero sets of initial conditions\cite{Suvakov&Dmitrasinovic13, Chenciner&Montgomery00}, there is no analytically tractable solution to the general, {\it non-hierarchical} three body problem.  Part of the reason for this is  the fundamentally chaotic nature of non-hierarchical triples, which causes astronomically slow convergence of perturbative solutions\cite{Sundman12}.  Absent a general analytic solution, much of our physical insight has instead come from numerical orbit integration\cite{Agekyan&Anosova67, HutBahcall83}.  These integrations demonstrate that non-hierarchical triples with negative total energy will generically disintegrate into a survivor binary and a single escaper star\cite{Standish72}.  This escape process sometimes occurs promptly, but sometimes takes many dynamical times to complete.  

In this paper, we complete the project initiated by Monaghan\cite{Monaghan76a}, and analytically compute the total accessible phase volume, $\sigma$, available to outcome states of the non-hierarchical three-body problem.  Unlike past attempts, our approach self-consistently accounts for both energy and angular momentum conservation, and quarantines the most uncertain assumptions (causality criteria) into a specific step of the computation, enabling future researchers - with, one can imagine, a greater understanding of the triple stability boundary - to improve the accuracy of our work.

Various approximations of Eq. \ref{eq:sigmaCart} have been used in the past to estimate the ``ergodic'' outcome distribution of the three-body problem\cite{Monaghan76a, Monaghan76b, Nash&Monaghan78, Valtonen+05, Valtonen&Karttunen06}.  This procedure assumes that, given a {\it microcanonical} ensemble of non-hierarchical three body systems with different initial conditions but otherwise identical integrals of motion and mass triplets, the outcome states (after breakup) will uniformly populate the accessible phase space volume.  This ensemble is microcanonical in the sense that each three-body system is isolated and not interacting with external sources of heat, but otherwise differs from the usual microcanonical ensemble in its very small particle number (similar approaches have a longer history of use in both nuclear\cite{Bohr36, Hauser&Feshbach52} and particle\cite{Fermi50} physics).  

The analytic and semi-analytic predictions of this statistical approach to the three-body problem generally fail to agree with detailed numerical results from three-body scattering simulations.  One possible explanation is the neglect of causality constraints in computations of the accessible phase space volume; by inserting an approximate version of these constraints into a simplified phase space volume calculation, some studies have obtained better agreement with numerical scattering simulations\cite{Valtonen+05} .  However, past attempts to include causality constraints were not truly first-principles calculation, as they neglected angular momentum conservation; furthermore, the current lack of analytic clarity on general triple stability criteria makes it hard to delineate exact causality conditions.

The phase volume $\sigma$ defined in Eq. \ref{eq:sigmaCart} uses relative coordinates between the components of the surviving binary $\{\vec{r}, \vec{p}\}$, and also relative coordinates between the escaping single star and the binary center of mass, $\{\vec{r}_{\rm s}, \vec{p}_{\rm s}\}$; the phase volume is evaluated at the moment of breakup, and the reference frame is in the binary center of mass.  Three clear assumptions enter into this formalism: that outcomes are uniformly distributed through the accessible phase space, that there is a well-defined moment of disintegration, and that at the time of disintegration, the trajectories can be decomposed into two pairwise orbits.  A fourth assumption enters implicitly, through the limits of integration: that there is a well-defined ``strong interaction region'' interior to which disintegration of the metastable triple may occur.  This fourth assumption is the most complicated, and we return to it in greater detail later.

By working in relative coordinates (i.e. treating the binary as a point mass when computing the hyperbolic trajectory of the escaper; neglecting the escaper's perturbations on the internal motion of the binary), the outcome phase space is 12-dimensional, but many portions of it are inaccessible due to conservation of energy and angular momentum.   Early efforts computed the ergodic outcome distribution by restricting the outcomes to an 11-dimensional hypersurface under the assumption of energy conservation; the neglect of angular momentum conservation was assumed to be appropriate for low angular momentum systems\cite{Monaghan76a}.  This approach was soon generalized to allow for angular momentum conservation in the special case where all motion is planar\cite{Monaghan76b}, although both of these works neglected the interaction energy between the escaper and the survivor binary (i.e. straight line escape trajectories).  Later, a general formalism was presented for estimating ergodic outcomes in the fully three-dimensional case, allowing for both energy and angular momentum conservation\cite{Nash&Monaghan78}.  However, the mathematical difficulty of the full problem prevented the calculation of closed-form outcome distributions, even neglecting interaction energy, and these results were evaluated numerically.  This formalism has also been extended to the Newtonian four-body problem\cite{Nash&Monaghan80}, with more limited results.

The angular momentum constraints on phase space volume $\sigma$ can be rewritten component-wise as
\begin{equation}
\delta(\vec{L}_{\rm B}+\vec{L}_{\rm s}-\vec{L}_{0})=\delta(L_{\rm B,x}+L_{\rm s,x})\delta(L_{\rm B,y}+L_{\rm s,y})\delta(L_{\rm B,z}+L_{\rm s,z}-L_0)
\end{equation}
if we limit degrees of freedom by picking a Cartesian coordinate system such that $\hat{z} \parallel \vec{L}_0$.  However, even with this convenient assumption, these integrals appear intractable in rectilinear coordinates\cite{Nash&Monaghan78}, and to make progress we shall switch to a more physically motivated coordinate system where angular momentum components have a simpler representation.  A tempting candidate would be Keplerian orbital elements, e.g. $\{\vec{r}, \vec{p}\} \rightarrow \vec{K}=\{a, e, I, \Omega, \omega, \lambda \}$.  These orbital elements represent semimajor axis, eccentricity, inclination, longitude of ascending node, longitude of pericenter, and mean anomaly, respectively.  However, the Jacobian for this transformation is sufficiently complicated that it is not even clear if this choice of coordinates would aid simplification of $\sigma$.

Instead, we will transform to Delaunay elements $\vec{D}$, an alternative parametrization of the two-body problem which has the virtue of being a canonical coordinate system.  The transformation $\{\vec{r}, \vec{p}\} \rightarrow \vec{D}=\{\Lambda, \Gamma, H, \eta, \omega, \lambda \}$ is therefore a symplectic one, with a Jacobian equal to $1$.  We define the elliptical Delaunay elements of the surviving binary in terms of standard Keplerian orbital elements as follows:
\begin{equation}
\begin{aligned}[l]
\Lambda &= \sqrt{Gm_{\rm B}a_{\rm B}} \\
\Gamma &=\sqrt{Gm_{\rm B}a_{\rm B}(1-e_{\rm B}^2)} \\
H &=\sqrt{Gm_{\rm B}a_{\rm B}(1-e_{\rm B}^2)}\cos I_{\rm B}
\end{aligned}
\qquad\qquad
\begin{aligned}[r]
\lambda&=\lambda_{\rm B} \\
\gamma&=\omega_{\rm B} \\
\eta&=\Omega_{\rm B}.
\end{aligned}
\end{equation}
While the canonical coordinates are simply the angular orbital elements from $\vec{K}$, the canonical momenta are different constants of the two-body problem.  We have placed a subscript ``B'' on the Keplerian elements to indicate their association with the survivor binary, and avoid confusion later on.  The Delaunay elements defined above are, strictly speaking, only valid for a bound orbit.  For the unbound orbit of the escaper, we will define its phase space position using hyperbolic\cite{Floria95} Delaunay elements $\vec{D}_{\rm H}$:
\begin{equation}
\begin{aligned}[l]
\mathcal{L} &= -\sqrt{GMa_{\rm s}} \\
\mathcal{G} &=\sqrt{GMa_{\rm s}(e_{\rm s}^2-1)} \\
\mathcal{H} &=\sqrt{GMa_{\rm s}(e_{\rm s}^2-1)}\cos I_{\rm s}
\end{aligned}
\qquad\qquad
\begin{aligned}[r]
\ell&=n_{\rm s}t \\
g&=\omega_{\rm s} \\
h&=\Omega_{\rm s}.
\end{aligned}
\end{equation}
Here we have used the hyperbolic Keplerian orbital elements (denoted with a subscript ``s'') for the unbound trajectory of the escaper star, and also its mean motion $n_{\rm s} = \sqrt{GM/a_{\rm s}^3}$.  Unlike all past approaches, our reparametrization of Eq. \ref{eq:sigmaCart} self-consistently accounts for the interaction energy between the escaper and the binary; the only approximation made is to treat the binary as a point particle.

Now we may begin simplifying the integrand of Eq. \ref{eq:sigmaCart} by rewriting constants of motion.  Specifically, we have
\begin{equation}
\begin{aligned}[l]
E_{\rm B}&=-\frac{G^2 m_{\rm a}m_{\rm b}m_{\rm B}}{2\Lambda^2} \\
L_{\rm B}&=\mathcal{M}\Gamma \\
L_{\rm B, z}&= \mathcal{M}H
\end{aligned}
\qquad\qquad
\begin{aligned}[r]
E_{\rm s}&=\frac{G^2 m_{\rm s}m_{\rm B}M}{2\mathcal{L}^2} \\
L_{\rm s}&=m\mathcal{G} \\
L_{\rm s, z}&=m\mathcal{H}.
\end{aligned}
\end{equation}
We note further that
\begin{equation}
\begin{aligned}[l]
L_{\rm B, x}&=L_{\rm B}\sin\eta\sin I_{\rm B} \\
L_{\rm B, y}&=-L_{\rm B}\cos\eta\sin I_{\rm B} 
\end{aligned}
\qquad\qquad
\begin{aligned}[r]
L_{\rm s, x}&=L_{\rm s}\sin h\sin I_{\rm s} \\
L_{\rm s, y}&=-L_{\rm s}\cos h\sin I_{\rm s},
\end{aligned}
\end{equation}
and that
\begin{equation}
\begin{aligned}[l]
\sin I_{\rm B} &= \sqrt{1-H^2/\Gamma^2}
\end{aligned}
\qquad\qquad
\begin{aligned}[r]
\sin I_{\rm s}=\sqrt{1-\mathcal{H}^2/\mathcal{G}^2}.
\end{aligned}
\end{equation}
Eq. \ref{eq:sigmaCart} can now be rewritten as 
\begin{align}
\sigma=& \idotsint \delta \left(\frac{G^2m_{\rm s}m_{\rm B}M}{2\mathcal{L}^2}-\frac{G^2m_{\rm a}m_{\rm b}m_{\rm B}}{2\Lambda^2}-E_0 \right) \label{eq:sigmaDel}\delta(\mathcal{M}H+m\mathcal{H}-L_0) \\
&\times \delta(\mathcal{M}\Gamma\sin\eta\sin I_{\rm B} + m\mathcal{G}\sin h \sin I_{\rm s}) \delta(\mathcal{M}\Gamma\cos\eta\sin I_{\rm B} + m\mathcal{G}\cos h \sin I_{\rm s}){\rm d}\vec{D}{\rm d}\vec{D}_{\rm H} . \notag
\end{align}
We begin trivially, by integrating ${\rm d}g$ and ${\rm d}\gamma$ from $0$ to $2\pi$.  The next step, which is to integrate ${\rm d}\ell$ and ${\rm d}\lambda$, appears just as simple; much like the longitudes of pericenter, the mean anomalies are absent from the integrand.  However, this step is a critical and conceptually subtle one, as it asks the question: in what sense is the outcome distribution ``ergodic?''  Do we consider different escapers from our hypothetical ensemble, viewed at fixed times $t$ post-ejection?  Do we consider them within a range of anomalies $\ell$?  Or do we consider them within a range of radii $r_{\rm s}$?  

Past examinations of the three-body problem chose the latter of these three options \cite{Monaghan76a, Valtonen&Karttunen06}, comparing phase volumes at a fixed $r_{\rm s}$ equal to a small multiple of $a_{\rm B}$.  Physically, we can understand this as an application of the ergodic hypothesis at the ``moment of breakup.''  The metastable triple is assumed to ergodically explore its accessible hypersurface until the precise moment of breakup, which is idealized as occurring at a fixed separation $r_{\rm s} \le R$ from the binary center of mass.  In principle, $R$ may be a function of many (perhaps all) of the Delaunay variables in this problem.  More recent statistical examinations of non-hierarchical triples defined a causal escape criterion in a highly simplified way\cite{Valtonen+05, Valtonen&Karttunen06}, with $R\equiv \alpha a_{\rm B}$, where $\alpha$ is a dimensionless number that can be calibrated from numerical scattering experiments.  For now, we will remain slightly more agnostic on the nature of triple breakup, and define the moment of breakup as occurring at an $r_{\rm s} \le R(\Lambda, \Gamma, H)$, with explicit functional forms for $R$ to be explored later.  The introduction of this idealization (and fudge factors such as $\alpha$) is unappealing, but as we shall see, it has a limited impact on the outcome distributions.

We have explored other invocations of the ergodic hypothesis, e.g. constant $t$ or constant $\ell$.  The first of these does not seem well-motivated to us, and yields outcome distributions very different from experiment.  For strongly hyperbolic escape, a constant $\ell$ is not too different from constant $r_{\rm s}$, but this similarity breaks down for nearly parabolic escapers, in a way that makes the outcome distribution ill-defined and divergent (a vice shared by the constant $t$ choice).  For the remainder of this paper, we assume that the metastable triple's motion is ergodic up until the point of breakup, which occurs within the interaction region $r_{\rm s}\le R$.  

We now consider our ensemble of escapers within a fixed range $0\le \ell \le \ell_{\rm max}$, where $\ell_{\rm max}$ corresponds to an orbital separation $r_{\rm s}=R$, at the edge of the interaction region.  For a hyperbolic trajectory, 
\begin{align}
&\ell_{\rm max} = \sqrt{\frac{R^2}{a_{\rm s}^2} + \frac{2R}{a_{\rm s}} + 1 - e_{\rm s}^2 } - \acosh\left(\frac{R/a_{\rm s} + 1}{e_{\rm s}} \right) \\
& = \sqrt{ \frac{G^2M^2R^2}{\mathcal{L}^4 } + \frac{2GMR}{\mathcal{L}^2} - \frac{\mathcal{G}^2}{\mathcal{L}^2}}  - \acosh \left(\frac{GMR /\mathcal{L}^2 + 1}{\sqrt{1+\mathcal{G}^2/\mathcal{L}^2} }\right). \notag
\end{align}
While the hyperbolic mean anomaly $\ell$ may range from 0 (ejection at pericenter) to $\ell_{\rm max}$ (ejection at the farthest point along the orbit inside the interaction region), the elliptical mean anomaly $\lambda$ only ranges across $\{0, 2\pi\}$; if $\lambda$ were permitted to grow without bound, the phase volume accessible to an elliptical orbit would diverge in time.  Integrating ${\rm d}\ell$ and ${\rm d}\lambda$, we find that the phase volume is now
\begin{align}
\sigma=&(2\pi)^3  \idotsint \delta\left(\frac{G^2m_{\rm s}m_{\rm B}M}{2\mathcal{L}^2}-\frac{G^2m_{\rm a}m_{\rm b}m_{\rm B}}{2\Lambda^2}-E_0 \right)  \delta(\mathcal{M}H+m\mathcal{H}-L_0)  \\
&\times \delta(\mathcal{M}\Gamma\sin\eta\sin I + m\mathcal{G}\sin h \sin I_{\rm s}) \delta(\mathcal{M}\Gamma\cos\eta\sin I + m\mathcal{G}\cos h \sin I_{\rm s})  \ell_{\rm max}   \notag \\
& \times {\rm d}\Lambda{\rm d}\Gamma{\rm d}H{\rm d}\eta{\rm d}\mathcal{L}{\rm d}\mathcal{G}{\rm d}\mathcal{H}{\rm d}h. \notag
\end{align}
For brevity we have used $\ell_{\rm max}(\mathcal{L}, \mathcal{G})$, rather than writing this term explicitly.

Having removed all four coordinates that do not appear in the integrand, we are left with eight variables.  Our goal is now to use the remaining integrals of motion to eliminate three canonical momenta and both nodal angles.  We shall integrate out the canonical momenta of the escaper, and integrate over any surviving nodal angles, to leave behind a three-variable probability density function in the integrand describing the distribution of binary parameters $\{\Lambda, \Gamma, H\}$. 

We proceed with a change of variables $\{\mathcal{G}, h, \mathcal{H} \} \rightarrow \{z_1, z_2, \mathcal{H} \} $, where $z_1= m \sin h \sqrt{\mathcal{G}^2-\mathcal{H}^2}$, and $z_2= m \cos h \sqrt{\mathcal{G}^2-\mathcal{H}^2}$.  Both $z_1$ and $z_2$ range from $-m | \mathcal{L}|$ to $m| \mathcal{L}|$.  We compute the Jacobian for this transformation by rewriting $\mathcal{G}=\sqrt{m^{-2}(z_1^2+z_2^2) + \mathcal{H}^2}$ and $h=\atan(z_1/z_2)$.  The resulting Jacobian determinant is $J_1 = m^{-1} (z_1^2+z_2^2+m^2 \mathcal{H}^2)^{-1/2}$, so we now have
\begin{align}
\sigma=&\frac{(2\pi)^3}{m}  \idotsint \delta\left(\frac{G^2m_{\rm s}m_{\rm B}M}{2\mathcal{L}^2}-\frac{G^2m_{\rm a}m_{\rm b}m_{\rm B}}{2\Lambda^2}-E_0 \right)  \delta(\mathcal{M}H+m\mathcal{H}-L_0) \notag \\
&\times\delta(z_1 + \mathcal{M}\sin\eta\sqrt{\Gamma^2-H^2}) \delta(z_2 + \mathcal{M}\cos\eta\sqrt{\Gamma^2-H^2}) \left(z_1^2+z_2^2+m^2\mathcal{H}^2 \right)^{-1/2} \notag \\
& \times \ell_{\rm max} {\rm d}\Lambda{\rm d}\Gamma{\rm d}H{\rm d}\eta{\rm d}\mathcal{L}{\rm d}\mathcal{H}{\rm d}z_1{\rm d}z_2.
\end{align}
Integrating ${\rm d}z_1$ and ${\rm d}z_2$ we find
\begin{align}
\sigma=&\frac{(2\pi)^3}{m}  \idotsint \delta\left(\frac{G^2m_{\rm s}m_{\rm B}M}{2\mathcal{L}^2}-\frac{G^2m_{\rm a}m_{\rm b}m_{\rm B}}{2\Lambda^2}-E_0\right) \delta(\mathcal{M}H+m\mathcal{H}-L_0) \notag \\
& \times \Bigg(\sqrt{ \frac{G^2M^2R^2}{\mathcal{L}^4 } + \frac{2GMR}{\mathcal{L}^2} - \frac{m^2\mathcal{H}^2 + \mathcal{M}^2(\Gamma^2-H^2)}{m^2\mathcal{L}^2}}  \notag \\
& -\acosh\left( \frac{1+GMR/\mathcal{L}^2}{\sqrt{1+m^{-2}\mathcal{L}^{-2}(m^2\mathcal{H}^2 + \mathcal{M}^2(\Gamma^2-H^2) )}} \right) \Bigg) \notag \\
& \times  \left(\mathcal{M}^2(\Gamma^2-H^2)+m^2\mathcal{H}^2 \right)^{-1/2} {\rm d}\Lambda{\rm d}\Gamma{\rm d}H{\rm d}\eta{\rm d}\mathcal{L}{\rm d}\mathcal{H}. 
\end{align}
In this process, we have eliminated the $\eta$-dependence of the integrand, which now integrates trivially from $0$ to $2\pi$.  We also perform a subsequent integral over ${\rm d}(m\mathcal{H})$, yielding
\begin{align}
\sigma=&\frac{(2\pi)^4}{m^2} \idotsint \delta\left(\frac{G^2m_{\rm s}m_{\rm B}M}{2\mathcal{L}^2}-\frac{G^2m_{\rm a}m_{\rm b}m_{\rm B}}{2\Lambda^2}-E_0 \right)   \frac{{\rm d}\Lambda{\rm d}\Gamma{\rm d}H{\rm d}\mathcal{L}}{(\mathcal{M}^2(\Gamma^2-H^2) + (\mathcal{M}H - L_0)^2)^{1/2}}   \notag \\
& \times \Bigg(\sqrt{ \frac{G^2M^2R^2}{\mathcal{L}^4 } + \frac{2GMR}{\mathcal{L}^2} - \frac{\mathcal{M}^2(\Gamma^2-H^2) + (\mathcal{M}H-L_0)^2}{m^2\mathcal{L}^2}} \notag \\
& -\acosh\left( \frac{1+GMR/\mathcal{L}^2}{\sqrt{1+m^{-2}\mathcal{L}^{-2}(\mathcal{M}^2(\Gamma^2-H^2) + (\mathcal{M}H - L_0)^2)}} \right) \Bigg)
\end{align}
Together, the ${\rm d}z_1$, ${\rm d}z_2$, and ${\rm d}(m\mathcal{H})$ integrals eliminated the three $\delta$-functions enforcing angular momentum conservation.  We eliminate the final variable of the escaper's motion by transforming to $y\equiv G^2m_{\rm s}m_{\rm B}M/(2\mathcal{L}^2)$, and integrating ${\rm d}y$, so that
\begin{align}
&\sigma=\frac{2^{5/2}\pi^4GM^{5/2}}{(m_{\rm s}m_{\rm B})^{3/2}} \idotsint y^{-3/2}\delta(y -\frac{G^2m_{\rm a}m_{\rm b}m_{\rm B}}{2\Lambda^2}-E_0)   \frac{{\rm d}\Lambda{\rm d}\Gamma{\rm d}H{\rm d}y}{(\mathcal{M}^2(\Gamma^2-H^2) + (\mathcal{M}H - L_0)^2)^{1/2}}  \notag \\
&\times \Bigg(\sqrt{ \frac{4y^2R^2}{G^2m_{\rm s}^2m_{\rm B}^2 } + \frac{4yR}{Gm_{\rm s}m_{\rm B}} - \frac{\mathcal{M}^2(\Gamma^2-H^2) + (\mathcal{M}H-L_0)^2}{G^2m_{\rm s}^3m_{\rm B}^3/(2My)}} \notag \\
& -\acosh\left( \frac{1+2yR/(Gm_{\rm s}m_{\rm B})}{\sqrt{1+2yM(\mathcal{M}^2(\Gamma^2-H^2) + (\mathcal{M}H - L_0)^2)/(G^2m_{\rm s}^3m_{\rm B}^3)}} \right) \Bigg) . 
\end{align}
Thus, the total accessible phase volume of a metastable triple, at the moment of breakup, can be reduced to the following triple integral over the three non-trivial Delaunay elements of the survivor binary:
\begin{align}
&\sigma=\frac{2^{5/2}\pi^4GM^{5/2}}{(m_{\rm s}m_{\rm B})^{3/2}} \iiint  \frac{(E_0 - E_{\rm B})^{-3/2}{\rm d}\Lambda{\rm d}\Gamma{\rm d}H}{(\mathcal{M}^2(\Gamma^2-H^2) + (\mathcal{M}H - L_0)^2)^{1/2}}\notag \\
& \Bigg(\sqrt{\frac{2M(E_0-E_{\rm B})}{G^2m_{\rm s}^3 m_{\rm B}^3} } \sqrt{ 2m(E_0 - E_{\rm B}) R^2 + 2GMm^2R - (\mathcal{M}^2(\Gamma^2-H^2) + (\mathcal{M}H-L_0)^2)} \notag \\
& -\acosh\left( \frac{1+2(E_0 - E_{\rm B})R/(Gm_{\rm s}m_{\rm B})}{\sqrt{1+2M(E_0 - E_{\rm B})(\mathcal{M}^2(\Gamma^2-H^2) + (\mathcal{M}H - L_0)^2)/(G^2m_{\rm s}^3m_{\rm B}^3)}} \right) \Bigg)  . \label{eq:solution}
\end{align}
For brevity, we have written $-G^2m_{\rm a}m_{\rm b}m_{\rm B}/(2\Lambda^2)$ as $E_{\rm B}$.  The triple integral in Eq. \ref{eq:solution} represents the total accessible phase volume, and its integrand, which we can label ${\rm d}\sigma/{\rm d}\Lambda{\rm d}\Gamma{\rm d}H$, is the trivariate, differential distribution of outcomes with respect to $\{ \Lambda, \Gamma, H\}$.  This integrand is thus the microcanonical ensemble for the non-trivial outcome variables of the non-hierarchical three-body problem.  There are other outcome variables in the ensemble (specifically, the Delaunay elements $\gamma$, $\eta$, $\lambda$) that are uniformly distributed between $0$ and $2\pi$, while $\ell$ is uniformly distributed along a range specified by $R(\Lambda, \Gamma, H)$.  The nodal angles $\eta$ and $h$ are confined to a one-dimensional manifold by the conjunction of angular momentum conservation and our choice of coordinate system $\hat{L}_0 \parallel \hat{z}$; this can be viewed as a uniform distribution of $\eta$ from $0$ to $2\pi$ with $h$ then specified deterministically.  The canonical momenta of the escaper ($\mathcal{L}$, $\mathcal{G}$, $\mathcal{H}$) were eliminated from this calculation in the same way the escaper's nodal angle $h$ was.  For a given combination of $\Lambda$, $\Gamma$, and $H$, conservation of the integrals of motion allow the escaper variables to be computed; likewise, Eq. \ref{eq:solution} could be recast in terms of the Delaunay elements of the escaper, rather than the survivor binary. 

\begin{figure*}
\centering
\includegraphics[width=110mm]{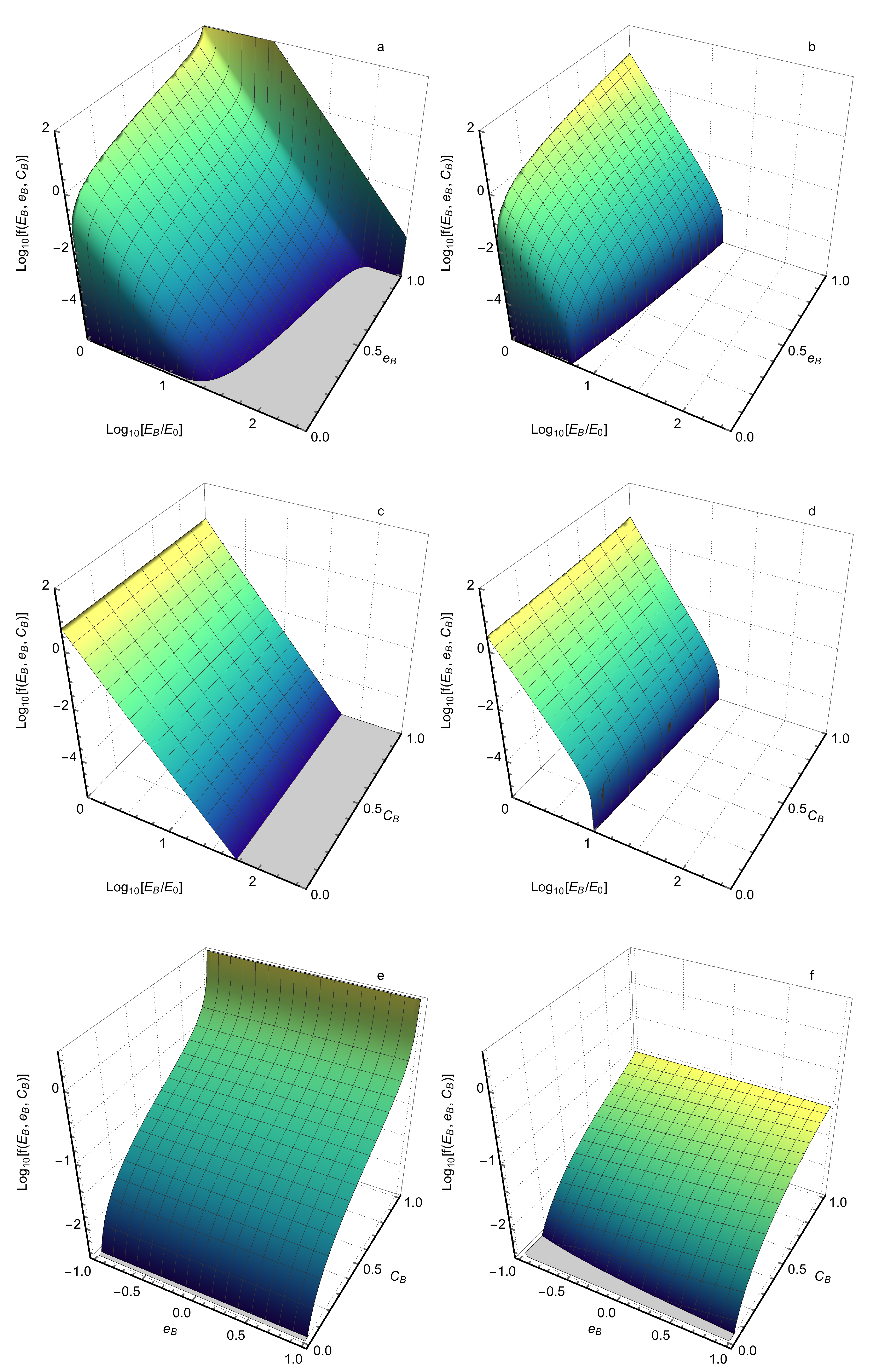}
\caption{{\bf The ergodic outcome distribution $f\equiv{\rm d}\sigma/{\rm d}E_{\rm B}{\rm d}e_{\rm B}{\rm d}C_{\rm B}$, as computed in Eq. \ref{eq:solution3}}.  Each plot shows a two-dimensional slice of the outcome distribution: in panels ({\bf a, b}), we fix $C_{\rm B}=0$; in panels ({\bf c, d}), fixed $e_{\rm B}=0.5$; in panels ({\bf e, f}), fixed $E_{\rm B}/E_0 = 0.5$.  Panels ({\bf a, c, e}) show $\tilde{L}_0=0$, while panels ({\bf b, d, f}) show $\tilde{L}_0=0.8$.}
\label{fig:phase3DTable}
\end{figure*}

The derivation of Eq. \ref{eq:solution} differs from past ergodic analyses of the non-hierarchical three-body problem by (i) conserving angular momentum, (ii) using a general (i.e. non-planar) geometry, (iii) accounting for the interaction potential between the binary and the escaper (no straight-line escape trajectories), (iv) producing a mathematically well-defined (non-divergent) estimate of phase volume, and (v) producing a closed-form expression for the distribution of outcomes.  Of the past efforts in this direction, some\cite{Monaghan76a, Valtonen+05} satisfied (ii, iv, v), one\cite{Monaghan76b} satisfied (i, v), and one\cite{Nash&Monaghan78} satisfied (i, ii), but none have previously satisfied all simultaneously.

One common uncertainty in all these works, and in ours, is the delicate question of how to define an ``interaction region'' inside of which the metastable triple may evolve ergodically.   Early efforts that neglected this issue incorporated a large phase volume of acausal escape trajectories; specifically, if one considers every position and velocity vector inside a sphere, one can find escape trajectories that do not time reverse into the binary.  This was somewhat rectified with the approximate inclusion\cite{Valtonen+05} of a ``loss cone'' into Eq. \ref{eq:sigmaCart}.  Our use of Delaunay elements accounts for causality in a way that is more accurate (unlike the straight-line escape trajectories implicit in the loss cone formalism\cite{Valtonen+05}, our escapers move along the full range of parabolic-to-hyperbolic orbits) {\it and} more transparent.  We have used the approximation of an interaction region of radius $R$, interior to which escapers must have their orbital pericenters.  In the remainder of this work, we consider two different escape criteria.  In analogy to the loss cone formalism\cite{Valtonen+05}, we consider a ``simple escape'' (SE) criterion, where $R \equiv \alpha a_{\rm B}$, but the larger geometric extent of highly eccentric binaries motivates us to consider an ``apocentric escape'' (AE) criterion as well, with $R \equiv \alpha a_{\rm B}(1+e_{\rm B})$.  While we certainly do not believe that these are exact models, they have the benefit of transparency: different escape criteria can be inserted into Eq. \ref{eq:solution} by varying the limits of integration and/or replacing our criteria with any more general $R(\Lambda, \Gamma, H)$.

In the next section, we take important physical limits of this joint distribution, rewrite it in terms of more familiar orbital properties, and otherwise interpret the physical outcomes of ergodic, non-hierarchical three-body encounters.

\section{Outcomes of Non-Hierarchical Three Body Encounters}
\label{sec:outcomes}
The Delaunay elements used in this derivation are slightly less intuitive than standard Keplerian orbital elements, so to gain more physical insight, we transform our outcome distribution into other sets of variables.  In particular, we consider the mappings $\{\Lambda, \Gamma, H\} \rightarrow \{E_{\rm B}, L_{\rm B}, C_{\rm B}\}$, and $\{\Lambda, \Gamma, H\} \rightarrow \{E_{\rm B}, e_{\rm B}, C_{\rm B}\}$, where $C_{\rm B}=\cos I$.  The Jacobian matrices for these transformations are triangular, so their determinants are simply $J_2=2^{-3/2}G\mathcal{M}^{-2}(m_{\rm a}m_{\rm b}m_{\rm B})^{1/2}L_{\rm B}(-E_{\rm B})^{-3/2}$ and $J_3=2^{-5/2}G^3e_{\rm B}(m_{\rm a}m_{\rm b}m_{\rm B})^{3/2}(-E_{\rm B})^{-5/2}$, respectively.  Applying these changes, we find Eq. \ref{eq:solution2} (using $J_2$), and, with $J_3$,
\begin{align}
\sigma&= \frac{\pi^4G^4 M^{5/2} (m_{\rm a} m_{\rm b})^{3/2} }{m_{\rm s}^{3/2}} \iiint \frac{e_{\rm B} {\rm d}E_{\rm B}{\rm d}e_{\rm B} {\rm d}C_{\rm B}}{L_{\rm s}(-E_{\rm B})^{5/2}(E_0 - E_{\rm B})^{3/2}} \notag \\
& \times \Bigg(\sqrt{\frac{2M(E_0-E_{\rm B})}{G^2m_{\rm s}^3 m_{\rm B}^3} } \sqrt{ 2m(E_0 - E_{\rm B}) R^2 + 2GMm^2R - L_{\rm s}^2} \notag \\
& -\acosh\left( \frac{1+2(E_0 - E_{\rm B})R/(Gm_{\rm s}m_{\rm B})}{\sqrt{1+2M(E_0 - E_{\rm B})L_{\rm s}^2/(G^2m_{\rm s}^3m_{\rm B}^3)}} \right) \Bigg).  \label{eq:solution3}
\end{align}
For brevity, we have re-inserted the angular momentum of the escaping star,
\begin{equation}
L_{\rm s}^2 \equiv L_{\rm B}^2(1-C_{\rm B}^2) + (L_{\rm B}C_{\rm B} - L_0)^2. \label{eq:surplus}
\end{equation}
Angular momentum conservation allows us to express $L_{\rm s}$ purely as a function of $L_{\rm B}$ and $C_{\rm B}$.

When $L_0=0$, the integrand of Eq. \ref{eq:solution3} can remain real-valued for arbitrarily large $|E_{\rm B}|$, but when $L_0>0$, there is an upper limit, $|E_{\rm max}|$, to the binding energy of the survivor binary.  The mathematical origin of this upper limit is quite clear: arbitrarily large binary energies would make both terms of the integrand imaginary-valued.  More specifically, at $E_{\rm B}=E_{\rm max}$, the argument of the large radical is equal to zero, and the argument of the $\acosh$ is equal to unity; when $|E_{\rm B}| > |E_{\rm max}|$, these arguments are less than zero and less than one, respectively, so their functions are imaginary-valued (notably, the exact same value of $E_{\rm B}$ is the critical $E_{\rm max}$ for each term).  Physically, this limit can be understood primarily through angular momentum conservation, which requires arbitrarily large velocities if all three bodies are confined to an arbitrarily small volume - as is necessary in our formalism to obtain arbitrarily large $|E_{\rm B}|$.  If, at the moment of disintegration, $|E_{\rm B}| \gg |E_0|$, it can become impossible to simultaneously satisfy the constraint that $r_{\rm s} \le R(E_{\rm B})$ along with conservation of energy and angular momentum.  The orbital separation $r_{\rm s}$ that satisfies all three of these conditions will be smaller than the pericenter of the hyperbolic orbit, and its true anomaly $\ell$ will thus be imaginary.

Under the approximation that $|E_0| \ll |E_{\rm B}|$, we solve exactly for this maximum energy.  For a given pair of $e_{\rm B}$ and $C_{\rm B}$ values, it is 
\begin{align}
&|E_{\rm max}'| = \frac{m_{\rm B} |E_0| }{m_{\rm a}^2 m_{\rm b}^2 \tilde{L}_0^2}  \Bigg( \mathcal{M}C_{\rm B}\sqrt{m_{\rm B}(1-e_{\rm B}^2)} \\
& + \sqrt{A^2m_{\rm a}m_{\rm b}m+2A m^2M-\mathcal{M}^2m_{\rm B}(1-C_{\rm B}^2)(1-e_{\rm B}^2)} \Bigg)^2. \notag
\end{align}
Here, we have used the symbol $A$ to represent $\alpha$ for the simple escape criterion, while it instead represents $\alpha(1+e_{\rm B})$ for apocentric escape.  For convenience, we have also introduced a dimensionless angular momentum $\tilde{L}_0 \equiv L_0 / L_{\rm c}(E_0)$, where the circular orbital angular momentum of a reference binary is $L_{\rm c}(E_0) \equiv G\mathcal{M}\sqrt{m_{\rm a}m_{\rm b} m_{\rm B} / (-2E_0)}$.  In the SE regime, we may compute a global maximum energy (across all values of $e_{\rm B}$ and $C_{\rm B}$), which is
\begin{equation}
|E_{\rm max}| =  \frac{m_{\rm B} |E_0| }{m_{\rm a}^2 m_{\rm b}^2 \tilde{L}_0^2}  \Bigg( \mathcal{M}\sqrt{m_{\rm B}}+ \sqrt{\alpha^2m_{\rm a}m_{\rm b}m+2\alpha m^2M} \Bigg)^2. 
\end{equation}
The global maximum energy in the AE regime is generally a factor $\approx 2$ larger, but cannot be expressed in a simple closed form.  Binary energies larger in magnitude than $|E_{\rm max}|$ are incompatible with a ``causal'' ejection, in that the pericenter of the ejected star would have to be outside the interaction radius $R$.  This cutoff in the energy distribution is one portion of our results sensitive to our idealizations concerning $R$, and is therefore worth special attention in future numerical scattering studies. 

The trivariate outcome distributions in Eqs. \ref{eq:solution2} and \ref{eq:solution} are somewhat complicated, so in Fig. \ref{fig:phase3DTable}, we plot two-dimensional slices of these distributions to aid in visualization.  In these figures, we show the outcome distribution as a function of two out of three of the integrand variables of Eq. \ref{eq:solution2}, while the third controlling variable has been set equal to a representative value.  We also plot the outcome distribution for two different values of total angular momentum, $\tilde{L}_0=0$ and $\tilde{L}_0=0.8$.  In two of the $\tilde{L}_0=0.8$ slices, the high-energy cutoff at $E_{\rm B}=E_{\rm max}$ is visible.  Other features of these distributions are explained in greater quantitative detail in the following subsections.

One important issue is that of convergence; superficially, it appears that the outcome distribution should diverge ($\propto (E_0 - E_{\rm B})^{-3/2}$) as $E_{\rm B} \to E_0$, and that the accessible phase volume is therefore unbounded.  This apparent divergence is regulated, however, by the radical and $\acosh$ terms in Eq. \ref{eq:solution}.  It is tedious but straightforward to apply L'H\^opital's rule to the integrand, and we find that
\begin{align}
\sigma_0 =& \lim_{E_B \to E_0}  \frac{{\rm d}\sigma}{{\rm d}E_{\rm B}{\rm d}L_{\rm B}{\rm d}C} = \frac{8\pi^4 M^{7/2}}{3G(m_{\rm a}m_{\rm b})^{3/2}m_{\rm s}^{11/2}m_{\rm B}^{3}} \label{eq:phaseLimit}  \\
& \times  \frac{L_{\rm B}}{(-E_0)^{3/2}L_{\rm s}} \frac{2G^2 m^3 M^2 R^2 + GMm L_{\rm s}^2 R - L_{\rm s}^4/m}{\sqrt{4Gm_{\rm s}m_{\rm B}R - 2L_{\rm s}^2/m} }.  \notag
\end{align}
The integrand is therefore convergent, and the phase volume is well-defined.

Should this be the case?  It is well-known that the phase volume accessible to the general $N$-body problem, on a hypersurface of constant energy, is divergent when $N>2$, rendering ergodic arguments ill-defined\cite{Padmanabhan90}.  We evade this problem by only considering metastable triples at the moment of breakup; if we followed the escaper as $t\to \infty$, then $\sigma$ would diverge.  At the moment of breakup, $\sigma \sim (r)^3(v)^3(r_{\rm s})^3(v_{\rm s})^3 \propto (R^3)(R^{-3/2})(R^3)(R^{-3/2})$.  This volume has an ``IR cutoff'' because $R \lesssim Gm_{\rm a}m_{\rm b}/|E_0|$, eliminating the usual phase volume divergence of star clusters (which occurs as a low-mass halo inflates to hold zero binding energy\cite{Tremaine+86}).  A ``UV cutoff'' is provided by $|E_{\rm max}|$ (i.e. angular momentum conservation) except in the special case where $L_0=0$.  Even in this special case, as $R\to 0$, $\sigma \to 0$ too. 

\subsection{Marginal Distribution of Energies}
\label{sec:marginalEnergy}
While Eqs. \ref{eq:solution2} and \ref{eq:solution} specify a joint distribution in three variables, we are also interested in marginal distributions.  The full trivariate outcome distribution is sufficiently complex that it cannot be integrated analytically to yield exact marginal distributions, but it is straightforward to integrate numerically.  We show the marginal distributions of binary energy $E_{\rm B}$, eccentricity $e_{\rm B}$, and orientation $C_{\rm B}$, in Extended Data Figs. \ref{fig:marginalEnergy}, \ref{fig:marginalEccentricity}, and \ref{fig:marginalInclination}, respectively.  In each of these we consider scattering ensembles with different total angular momentum $\tilde{L}_0$, different combinations of masses $\{m_{\rm a}, m_{\rm b}, m_{\rm s}\}$, varied $\alpha$ parameters, and both of the escape criteria discussed earlier.  

We will begin by investigating the marginal distribution of binary energies, ${\rm d}\sigma / {\rm d}E_{\rm B}$.  Eq. \ref{eq:phaseLimit} has already shown us that this distribution takes on a finite value as $E_{\rm B} \to E_0$.  Conversely, $E_{\rm B}$ can only become infinitely large in the limit of $L_0 = 0$: for finite $L_0$, ${\rm d}\sigma / {\rm d}E_{\rm B}$ must roll over to zero for $|E_{\rm B}| \ge |E_{\rm max}|$.  The exact behavior of ${\rm d}\sigma / {\rm d}E_{\rm B}$ very close to these limits is complex, but in the large intermediate region it may be approximated quite simply, as the $\ell_{\rm max}$ term in the trivariate outcome distribution is almost constant away from these two boundaries.  If we approximate $\ell_{\rm max}$ as a constant, then for {\it either} simple or apocentric escape criteria, we find that 
\begin{equation}
\frac{{\rm d}\sigma }{ {\rm d}E_{\rm B}} \propto \frac{|E_{\rm B}|^{-5/2}}{L_{\rm s}(E_0 - E_{\rm B})^{3/2}}. \label{eq:energyDistribution}
\end{equation}
Generally, $|E_{\rm max}| \gg |E_0|$, so that the large intermediate range of outcome energies has a simple power-law distribution in energies, with a power-law index set by $\tilde{L}_0$.  When $\tilde{L}_0 \approx 0$, $L_{\rm s} \approx L_{\rm B} \propto |E_{\rm B}|^{-1/2}$.  Conversely, when $\tilde{L}_0 \approx 1$, $L_{\rm s} \approx L_0$ over a large region of phase volume.  Therefore, the ergodic energy distribution in a $\tilde{L}_0 \ll 1$ ensemble is roughly ${\rm d}\sigma / {\rm d}E_{\rm B} \propto |E_{\rm B}|^{-7/2}$, while in a $\tilde{L}_0\sim 1$ ensemble, the steeper ${\rm d}\sigma / {\rm d}E_{\rm B} \propto |E_{\rm B}|^{-4}$ is a good approximation.  

We numerically integrate over $L_{\rm B}$ and $C_{\rm B}$ in Extended Data Fig. \ref{fig:marginalEnergy} to show the exact marginal distribution of energies, ${\rm d}\sigma/{\rm d}E_{\rm B}$.  As is predicted by conservation of angular momentum, it is impossible to reach arbitrarily large $|E_{\rm B}|$ except in the $L_0 \to 0$ limit.  The precise location of $|E_{\rm max}|$ in energy space depends on the mass ratio of the problem (low-mass escapers, with $m_{\rm s} \ll m_{\rm B}$, are unable to carry large quantities of angular momentum out of the system, and thus yield a smaller $|E_{\rm max}|$ value), on the escape criterion used (SE vs AE), and on the value of the $\alpha$ parameter.  However, the slope of the energy distribution power-law at intermediate energies is independent of all these assumptions and parameter choices, and appears to always be in good agreement with the approximate power laws predicted by Eq. \ref{eq:energyDistribution}.

The quasi-power-law distributions of survivor binaries we predict are similar to the ${\rm d}\sigma / {\rm d}E_{\rm B}  \propto |E_{\rm B}|^{-9/2}$ distribution predicted by detailed balance arguments\cite{Heggie75}, but are somewhat shallower (notably, a Monaghan-type calculation that neglects angular momentum conservation but accounts for causality\cite{Valtonen+05} will also produce ${\rm d}\sigma / {\rm d}E_{\rm B} \propto |E_{\rm B}|^{-9/2}$).  Therefore, one prediction of our formalism is that a population of binaries evolving through a sequence of strong, chaotic three-body interactions will never achieve detailed balance: a net flow always exists from the ``softer'' (low $|E_{\rm B}|$) to the ``harder'' (high $|E_{\rm B}|$) end of the distribution.

\subsection{Marginal Distribution of Eccentricities}
\label{sec:marginalEccentricity}

The distribution of eccentricities is very important for understanding how exotic compact object binaries are dynamically produced through three-body scatterings in dense stellar environments.  In order to gain physical intuition, we will estimate the marginal distribution of eccentricity, ${\rm d}\sigma/ {\rm d}e_{\rm B}$, by approximating the $\ell_{\rm max}$ term (i.e. the difference of the radical and $\rm acosh$ terms) in Eq. \ref{eq:solution2}.  For comparable-mass systems, the first term ($\propto R^2$) inside the radical is generally the dominant component of $\ell_{\rm max}$, except in parts of phase volume very close to the boundary of the accessible hypersurface (e.g. the border where energy and angular momentum conservation become impossible to maintain).  We therefore approximate ${\rm d}\sigma / {\rm d}E_{\rm B}{\rm d}e_{\rm B}{\rm d}C_{\rm B} \propto e_{\rm B}L_{\rm s}^{-1}A$, where, as before, $A=\alpha$ for the simple escape criterion and $A=\alpha(1+e_{\rm B})$ for our fiducial model of apocentric escape.  In the $\tilde{L}_0 \approx 1$ limit, we again approximate $L_{\rm s} \approx L_0$, and find
\begin{align}
\frac{{\rm d} \sigma}{{\rm d} e_{\rm B}} \approx 
\begin{cases}
2e_{\rm B} & {\rm [SE]} \label{eq:marginalEccApproximate1} \\
\frac{6}{5}e_{\rm B}(1+e_{\rm B}) & {\rm [AE]}.
\end{cases}
\end{align}
For the simple escape criterion, the ergodic distribution of survivor eccentricities is, in the large-$L_0$ limit, {\it exactly thermal}.  In contrast, the apocentric escape criterion yields a mildly super-thermal outcome distribution due to the larger average interaction cross-section of a high-$e_{\rm B}$ binary, the apocenter of which is twice as large as that of a circular binary of equal energy.  In the $\tilde{L}_0 \approx 0$ limit, we will again approximate $L_{\rm s} \approx L_{\rm B} \propto \sqrt{1-e_{\rm B}^2}$, implying that
\begin{align}
\frac{{\rm d} \sigma}{{\rm d} e_{\rm B}} \approx 
\begin{cases}
e_{\rm B}/\sqrt{1-e_{\rm B}^2} & {\rm [SE]} \label{eq:marginalEccApproximate2} \\
\frac{4}{4+\pi}e_{\rm B}(1+e_{\rm B})/\sqrt{1-e_{\rm B}^2} & {\rm [AE]}.
\end{cases}
\end{align}
In the $\tilde{L}_0 \ll 1$ limit, the ergodic distribution of survivor eccentricities is thus {\it highly super-thermal}.  High-$e_{\rm B}$ binaries are strongly overproduced relative to a thermal eccentricity distribution, and indeed, ${\rm d}\sigma / {\rm d}e_{\rm B}$ possesses a removable singularity as $e_{\rm B} \to 1$.  This can be understood as a consequence of {\it angular momentum starvation}.  In a low-$L_0$ ensemble of non-hierarchical three-body systems, production of a $e_{\rm B} \approx 0$ survivor binary usually requires dramatic fine-tuning in the orientation of the escaper: it must leave on a retrograde trajectory, with a large pericenter and velocity, so as to carry away as much negative angular momentum as possible.  While this outcome is not generically prohibited, the degree of fine-tuning involved severely limits the accessible phase volume and disfavors its realization in an ergodic outcome distribution.

The approximate eccentricity distributions presented in Eqs. \ref{eq:marginalEccApproximate1} and \ref{eq:marginalEccApproximate2} were each derived under the assumption that the $R^2$ term dominates in $\ell_{\rm max}$.  This assumption is generally valid for comparable mass ratios, but breaks down when mass ratios are highly unequal.  When the escaper mass $m_{\rm s} \ll m_{\rm B}$, angular momentum starvation will again sculpt the outcome distribution into a highly nonthermal form: high-$L_0$ systems will exhibit a circular orbit bias, while low-$L_0$ systems can be even more radially biased than Eq. \ref{eq:marginalEccApproximate2} would predict.

In Extended Data Fig. \ref{fig:marginalEccentricity}, we compare these simple approximations to exact numerical evaluation of the marginal eccentricity distribution.  For $\tilde{L}_0 = 0$ and $\tilde{L}_0 = 1$ ensembles, Eqs. \ref{eq:marginalEccApproximate1} and \ref{eq:marginalEccApproximate2} provide excellent approximations in the comparable mass ratio regime.  These expressions only break down noticeably for $m_{\rm s} \lesssim 0.1 m_{\rm B}$.  In the comparable mass ratio regime, $\tilde{L}_0 = 0.5$ yields an outcome distribution closer to the high-$L_0$ limit.  The super-thermal outcome distributions we predict for low-$\tilde{L}_0$ systems have important implications for the dynamical formation of gravitational wave sources and accreting compact object binaries; we return to this topic in \S \ref{sec:disc}.

\subsection{Marginal Distribution of Inclinations}
\label{sec:marginalInclination}
As above, we can obtain useful physical intuition by dropping terms in the integrand of Eq. \ref{eq:solution2} to yield an approximate marginal distribution of binary inclinations, ${\rm d}\sigma/{\rm d}C_{\rm B}$.  We begin by dropping the radical and ${\rm acosh}$ terms, and integrating ${\rm d}L_{\rm B}$, from $L_{\rm B}=0$ to $L_{\rm B} = L_{\rm c}(E_{\rm B})$.  If we then approximate the resulting integrand by taking $E_{\rm B} \to E_0$ (crude simplifications, motivated by the steep decline in phase volume accessible at higher energies), we find 
\begin{align}
\frac{{\rm d}\sigma}{{\rm d}C_{\rm B}} \approx & k_{\rm C} \ln \left( \frac{L_{\rm c}-L_0 C_{\rm B} + \sqrt{L_{\rm c}^2 - 2L_0 L_{\rm c}C_{\rm B}+L_0^2} }{L_0 (1-C_{\rm B})} \right)\notag \\
& = k_{\rm C} \ln \left(1 + \frac{L_{\rm c} + L_{\rm s}(L_{\rm c}) - L_0}{L_0 (1-C_{\rm B})} \right), \label{eq:marginalC}
\end{align}
where $L_{\rm c} = G\mathcal{M}\sqrt{m_{\rm a} m_{\rm b} m_{\rm B} / 2|E_0|}$, $L_{\rm s}(L_{\rm c})$ is the escaper angular momentum when $L_{\rm B}=L_{\rm c}$, and $k_{\rm c}$ is a normalization constant defined such that $\int_{-1}^{+1} ({\rm d}\sigma/{\rm d}C_{\rm B})  {\rm d}C_{\rm B} = 1$.  Notable features of Eq. \ref{eq:marginalC} include a removable singularity as $C_{\rm B} \to 1$ (the integral of the distribution remains finite), and increasingly isotropic behavior as $L_0 \to 0$.  However, the distribution is ill-defined in the zero-angular momentum limit (where ${\rm d}\sigma/{\rm d}C_{\rm B}$ must be constant), and does not agree well with numerical evaluation of Eq. \ref{eq:solution2} for $\tilde{L}_0 \ll 0.1$.

We show a variety of marginal inclinations in Extended Data Fig. \ref{fig:marginalInclination}, where the distribution of survivor binary orientations are plotted against $C_{\rm B}$.  Eq. \ref{eq:solution2} predicts an isotropic distribution when $\tilde{L}_0=0$, as it must by symmetry.  Higher values of $\tilde{L}_0$ show a marked preference for prograde orbits, however.  The prograde bias of ${\rm d}\sigma / {\rm d}C_{\rm B}$ can be understood in terms of the bivariate distribution ${\rm d}\sigma/{\rm d}C_{\rm B}{\rm d}L_{\rm B}$, which we have marginalized over in our approximate derivation of Eq. \ref{eq:marginalC}.  This bivariate distribution scales $\propto L_{\rm s}^{-1}$, so the greatest phase volume exists when the survivor binary can ``soak up'' a large majority of $L_0$, which requires approximate alignment between $\vec{L}_{\rm B}$ and $\vec{L}_0$.

While Eq. \ref{eq:marginalC} provides an excellent approximation to the ergodic ${\rm d}\sigma/{\rm d}C_{\rm B}$ in the comparable-mass regime, regardless of escape criterion, it breaks down noticeably for low-mass escapers ($m_{\rm s} \lesssim m_{\rm a}/3$, for an equal-mass survivor binary).  The numerically evaluated marginal distributions we present in the unequal-mass regime deviate from Eq. \ref{eq:marginalC} because of angular momentum starvation: a much greater prograde bias sets in when the escaper is unable to carry away a large fraction of $L_0$ on a causal escape trajectory.

\section{Comparison to Numerical Scattering Experiments}
\label{sec:numerics}
In this section, we numerically integrate several ensembles of non-hierarchical three-body systems to test the predictions of \S \ref{sec:outcomes}.  
There are many analytic and semi-analytic predictions of the previous section, and it would be beyond the scope of this paper to fully map the parameter space of the three-body problem via numerical scattering experiments.  This section is instead a preliminary exploration into the ergodicity of non-hierarchical triples.  We focus our effort on three key predictions of Eq. \ref{eq:solution}: the marginal outcome distributions ${\rm d}\sigma/{\rm d}E_{\rm B}$, ${\rm d}\sigma/{\rm d}e_{\rm B}$, and ${\rm d}\sigma/{\rm d}C_{\rm B}$.

We calculate the outcomes of a series of single-binary (1+2) interactions using the \texttt{FEWBODY} numerical scattering code (the source code can be found at http://fewbody.sourceforge.net).  This code integrates the usual $N$-body equations in position-space in order to advance the system forward in time\cite{Fregeau+04}.  This is accomplished via the eighth-order Runge-Kutta Prince-Dormand integration method with adaptive time-stepping and ninth-order error estimation.

For all simulations, all objects are assumed to be point-particles (i.e., the radii are set to zero) of equal mass. All binaries have initial semi-major axes of 1 AU, and initial eccentricities $e_0$ as provided in Extended Data Table~1.  We set the impact parameter to zero and the initial relative velocity at infinity $v_{\rm rel}$ to $0.01v_{\rm crit}$.  Here, $v_{\rm crit}$ is the critical velocity, defined as the relative velocity at infinity required for a total encounter energy of zero.  The motivation for these choices is that, as found in previous studies\cite{Leigh+16a,Leigh+18}, lower relative velocities at infinity and smaller impact parameters maximize the probability of long-lived resonant interactions occurring, which is probably needed to uphold the assumption of ergodicity.  All angles defining the relative configurations of the binary orbital planes and phases are sampled to ensure isotropic scattering.  The number of simulations performed for each combination of initial conditions are provided in the second column of Extended Data Table~1.  We use standard criteria\cite{Fregeau+04} to determine when each integration is terminated, and adopt a tidal tolerance parameter of $\delta =10^{-5}$ for all simulations.

Many of our simulations do not result in a long-lived, resonant three-body system, but instead resolve promptly (either in an exchange or a flyby).  The underpinning of our analytic formalism is the ergodic hypothesis, and it is dubious that this principle would apply to non-resonant encounters, the behavior of which can be approximately analyzed with the impulse approximation (for the small impact parameters we are focused on here\cite{Heggie75, Heggie&Hut93}) and secular theory (for wider impact parameters\cite{Heggie75, Heggie&Rasio96, Hamers&Samsing19}).  We hypothesize that the key dynamical phase responsible for ``ergodicizing'' the outcomes of non-hierarchical triples is the ``scramble:'' a period of time when no pairwise binary exists.  This situation generally accounts for only a small minority of the lifetime of a metastable triple, reflecting the intermittently chaotic nature of these systems.  The typical metastable triple spends the bulk of its life evolving in a quasi-regular way during non-terminal ejections of single components, but when the ejected component returns to pericenter, it has the chance to enter a phase of intense chaos as it interacts strongly with both companions.  These periods of chaos are scrambles, and we count their number, $N_{\rm scram}$, for each integration in every ensemble.  In computing $N_{\rm scram}$, we exclude the initial scramble that is produced by default in every run as a result of our zero-impact parameter initial conditions.

The topological maps in Fig. \ref{fig:ergodicDevelopment} illustrate the progressive disappearance of regularity in an ensemble (Run A) of three-body integrations with increasing scramble count.  Each panel shows topological maps in {\it outcome space}, specifically, the space of surviving binary eccentricity $e_{\rm B}$ and orientation $C_{\rm B}$ (left column) and the space of surviving binary energy $E_{\rm B}$ and $C_{\rm B}$ (right column).  Different rows correspond to subsets of Run A.  The top row shows the full sample.  The middle row shows the subset with $N_{\rm scram}\ge 1$.  This row excludes all prompt exchanges and contains almost all resonant encounters.  The bottom row shows the subset of integrations with $N_{\rm scram}\ge 2$.  While the uppermost row has clear geometrical features indicative of regular evolution, these ``clouds of regularity'' disappear rapidly as one moves down.  These clouds represent a 2-manifold of regular outcomes living in the 3-dimensional outcome space; Fig. \ref{fig:ergodicDevelopment} illustrates their projection.  It seems that two or more actual scrambles usually suffice to remove memory of initial conditions.

This qualitative argument motivates our comparisons between the analytic formalism of Eq. \ref{eq:solution2} and ensembles of \texttt{FEWBODY} scattering experiments.  These comparisons, which are illustrated in Figs. \ref{fig:marginalEnergyNumeric} and \ref{fig:marginalMultiNumeric}, nvolve marginal distributions computed numerically from the trivariate outcome distribution in Eq. \ref{eq:solution2}.  In each case, we use the AE definition of the strong interaction region (as we shall see, our scattering experiments are in much better agreement with the AE than the SE hypothesis), and set $\alpha=2$.  The value of $\alpha$ is not strongly constrained by our scattering experiments, and a broad range of values are compatible with our results.  In both comparison figures, we use Poisson statistics\cite{Gehrels86} to compute $95\%$ confidence intervals for the marginal distribution in a given bin of $E_{\rm B}$, $e_{\rm B}$, or $C_{\rm B}$ (horizontal error bars are merely bin sizes).

In Fig. \ref{fig:marginalEnergyNumeric}, we perform this comparison for the marginal distribution of binary energies, ${\rm d}\sigma / {\rm d}E_{\rm B}$.  Runs A, B, and C are compared to the semi-analytic predictions (i.e. numerical marginalization) of Eq. \ref{eq:solution2}.  In the top panel of this figure, which examines the entirety of Runs A, B, and C, we see only approximate agreement between the ergodic theory and numerical experiment.  However, in the middle and bottom panels of this figure, we make increasingly restrictive cuts on the number of scrambles, $N_{\rm scram}$.  The bottom panel of Fig. \ref{fig:marginalEnergyNumeric}, showing the subsample of runs with $N_{\rm scram} \ge 2$, shows robust agreement between ergodic theory and experiment for the high-$L_0$ ensembles.  Run C, the low-$L_0$ ensemble ($\tilde{L}_0 = 0.44$), shows some disagreement in the high-energy tail of ${\rm d}\sigma / {\rm d}E_{\rm B}$: numerical scattering experiments seem to produce somewhat greater numbers of high-energy outcomes.  The extremely steep slope of the energy distribution means that we lack sufficient numerical resolution to strongly constrain the $\alpha$ parameter (although values of $\alpha \lesssim 1$ would produce visible disagreement).

In Fig. \ref{fig:marginalMultiNumeric}, we repeat this comparison for the marginal distribution of binary eccentricities, ${\rm d}\sigma/{\rm d}e_{\rm B}$.  As with the energy distribution, we see notable disagreement between the ergodic theory and numerical experiment in the top panel of this comparison, where we examine the entire scattering ensemble.  However, if we limit our comparison to the $\sim 50\%$ of the scattering experiments with $N_{\rm scram}\ge 2$, we find very good agreement between theory and experiment in Runs A and B, and reasonable agreement in the low-$L_0$ Run C.  As we predicted in \S \ref{sec:marginalEccentricity}, all scattering ensembles produce a super-thermal distribution of outcome eccentricities.  In high angular momentum ensembles, the distribution is mildly super-thermal, in excellent agreement with our ``apocentric escape'' hypothesis, but in notable disagreement with the (thermal) predictions of the SE hypothesis.  This implies that the larger geometric cross-section of a highly eccentric binary favors its production.  In Run C, we see evidence for an even greater bias towards low-$e_{\rm B}$ survivor binaries.  In the ergodic theory of \S \ref{sec:marginalEccentricity}, this greater degree of super-thermality is produced by angular momentum starvation.

In the right column of Fig. \ref{fig:marginalMultiNumeric}, we again repeat this comparison, this time for the marginal distribution of binary inclinations, ${\rm d}\sigma/{\rm d}C_{\rm B}$.  For Runs A, B, and C, the entire sample of scattering outcomes is in decent agreement with the ergodic theory, {\it except} for $C_{\rm B} \lesssim -0.8$, where scattering experiments produce a large tail of highly retrograde survivor binaries not predicted in \S \ref{sec:marginalInclination}.  As we have previously seen for ${\rm d}\sigma/{\rm d}E_{\rm B}$ and ${\rm d}\sigma/{\rm d}e_{\rm B}$, this disagreement subsides when we restrict ourselves to the subsamples with two or more scrambles.  In the bottom panel of Fig. \ref{fig:marginalMultiNumeric}, we see excellent agreement between all three scattering ensembles and their corresponding ergodic predictions.  The one exception to this agreement is, again, in Run C, where a residual tail of retrograde bias survives after the $N_{\rm scram} \ge 2$ cut.  We suspect that this small deviation may be due to the known tendency of Runge-Kutta scattering codes to produce integration errors during very close encounters (which are produced much more frequently in low-$L_0$ ensembles), and believe that this comparison should be re-examined in the future with a regularized algorithm, such as ARchain\cite{Mikkola&Aarseth90}.  As was predicted in \S \ref{sec:marginalInclination}, the tail of prograde equatorial outcomes is larger in lower-$L_0$ systems due to angular momentum starvation.

It is also worth comparing our predictions to previous power-law fits to numerical scattering experiments.  Detailed numerical scattering studies\cite{Valtonen+05} have fit power laws to two of the marginal distributions we have examined here.  Specifically,
\begin{enumerate}
\item The binary energy distribution was fit as ${\rm d}\sigma / {\rm d}E_{\rm B} \propto |E_{\rm B}|^{-n}$, where $n = 3+18\bar{L}_0^2$, $\bar{L}_0 = L_0 (2.5G m_0^{5/2} |E_0|)^{-1}$, and $m_0 = \sqrt{m_{\rm a}m_{\rm b} + m_{\rm a} m_{\rm s} + m_{\rm b} m_{\rm s}}/\sqrt{3}$.  For the equal-mass scattering ensembles considered in both these studies, $\bar{L}_0 = 0.32 \tilde{L}_0$.  The predicted power law indices for Runs A, B, and C are thus $n \approx 4.87$, $n\approx 4.42$, and $n\approx 3.36$, respectively.  These fitted power law indices are not too far from either the predictions of \S \ref{sec:outcomes} or our own scattering results in this section, although we note that in the chaotic ($N_{\rm scram}\ge 2$) scattering subsamples, we find power law indices in better agreement with \S \ref{sec:outcomes} than with these fitting formulas.  We speculate that the steeper slope of the fitting formula is due to the inclusion of promptly resolved encounters, as is suggested by an examination of Fig. \ref{fig:marginalEnergyNumeric}.
\item The binary eccentricity distribution was fit as ${\rm d}\sigma / {\rm d}e_{\rm B} \propto e (1-e^2)^p$, where $2p = \bar{L}_0 - 1/2$.  The predicted power law indices for Runs A, B, and C are $p=-0.089$, $p=-0.11$, and $p=-0.18$, respectively.  These fitted distributions are in reasonable agreement with our scattering results, although the functional form of the fit does not asymptote to our exact computation (${\rm d}\sigma / {\rm d}e_{\rm B} \propto e (1-e^2)^{-1/2}$) in the $L_0 \to 0$ limit.
\end{enumerate}

\section{Discussion}
\label{sec:disc}
We have completed the line of research initiated by Monaghan\cite{Monaghan76a} by deriving the ergodic distribution of outcomes for the chaotic 3-body problem.  More specifically, we have computed the eight-dimensional phase volume, $\sigma$, accessible to a microcanonical ensemble of non-hierarchical triples at the moment of disintegration.  This phase volume is represented (in Eqs. \ref{eq:solution2}, \ref{eq:solution}, \ref{eq:solution3}) as a triple integral, the integrand of which is the trivariate outcome distribution for the surviving binary system.  This outcome distribution can be written in closed form, and is most simply expressed in terms of binary energy $E_{\rm B}$, binary angular momentum $L_{\rm B}$, and the cosine of its orbital inclination (with respect to the triple's total angular momentum), $C_{\rm B}$.  We have computed marginal distributions, such as ${\rm d}\sigma/{\rm d}E_{\rm B}$, both numerically and in approximate analytic form.  

The primary conceptual assumptions in this calculation are three.  First, we assume that the chaotic evolution of the metastable triple is ergodic in nature, so that by the time of breakup, it is equally likely to exist at any point in its eight-dimensional phase space (the eighteen dimensions accessible to a system of three particles are reduced to twelve by shifting to relative positions, and then reduced to eight by integrals of motion).  Such ergodicity is not present in all chaotic systems, and islands of regularity are known to exist for the three-body problem, but this nonetheless seems like a reasonable starting point.  Second, we assume that the disintegration of the system can be approximated as instantaneous.  Third, we have implicitly assumed (by employing relative coordinates) that the escaping single star sees the receding binary roughly as a point particle.

One additional assumptions enters for practical purposes.  We assume that the triple is only able to break up and eject a single star within a certain region of strong interaction.  We parametrize this interaction region as a sphere of radius $R=\alpha a_{\rm B}(1+e_{\rm B})$, where the dimensionless number $\alpha \sim 1$.  This criterion is likely not exact, but again seems to be a reasonable approximation.  Fortunately, many of our results depend only weakly on $\alpha$, which is the sole free parameter in our formalism, and can in the future be better calibrated from larger ensembles of numerical scattering experiments.  We have also quarantined this assumption into the final step of our calculation, so that more detailed forms of the triple stability boundary\cite{Mardling&Aarseth01} can be easily inserted into our formalism. 

Even with these simplifying assumptions, the analytic computation of the ergodic outcome distribution is nontrivial, and past attempts have had to make other approximations of greater significance\cite{Monaghan76a, Monaghan76b, Nash&Monaghan78, Valtonen+05}, producing inaccurate results (see the discussion under Eq. \ref{eq:solution} for a comparison to past work).  We have made this problem analytically tractable with the use of canonical transformations that greatly reduce its difficulty.

Although the primary focus of this paper was the computation of the three-body problem's ergodic outcome distribution, we have also begun to compare our formalism's predictions to ensembles of numerically integrated three-body systems.  A fuller comparison will be the subject of future work, but we list here our primary predictions and a preliminary assessment of their correspondence to numerical scattering experiments.  In all these comparisons, we have focused on the $\sim 50\%$ of our numerical scattering experiments which pass through two or more ``scrambles,'' periods of dynamical activity where no pairwise binaries exist in the metastable triple.  We also focus here on comparable-mass systems, although our predictions for more extreme mass ratios are detailed in \S \ref{sec:outcomes}.
\begin{itemize}
\item The marginal distribution of binary energies, ${\rm d}\sigma/ {\rm d}E_{\rm B}$, is to a good approximation a power law $\propto |E_{\rm B}|^{-4}$ between the range $E_0$ and a maximum $E_{\rm max} \propto (\alpha/L_0)^{2}$.  In very low-angular momentum ensembles, the power law softens slightly (${\rm d}\sigma/ {\rm d}E_{\rm B} \propto |E_{\rm B}|^{-7/2}$), and in zero-angular momentum ensembles, $E_{\rm max}\to -\infty$.  Binaries more energetic than $E_{\rm max}$ cannot be produced due to angular momentum conservation, and the location of this ``UV cutoff'' is the only feature of the energy distribution sensitive to the fudge factor $\alpha$.  Our numerical scattering experiments confirm the power law behavior of ${\rm d}\sigma/{\rm d}E_{\rm B}$, but lack the resolution to measure $E_{\rm max}$.  The ergodic power law is close to but differs from that predicted from detailed balance considerations\cite{Heggie75} (${\rm d}\sigma / {\rm d}E_{\rm B} \propto |E_{\rm B}|^{-9/2}$), indicating that a population of binary stars evolving through a sequence of chaotic binary-single scatterings will be somewhat out of detailed balance.
\item The marginal distribution of binary eccentricities, ${\rm d}\sigma/{\rm d} e_{\rm B}$, varies between two asymptotic limits.  In an ensemble of high-angular momentum encounters ($\tilde{L}_0 \approx 1$), the ergodic outcome distribution is slightly super-thermal (biased towards an excess of highly eccentric orbits, due to the larger geometric cross-section of highly eccentric binaries), with ${\rm d}\sigma / {\rm d} e_{\rm B} \propto e_{\rm B}(1+e_{\rm B})$.  As the ensemble $\tilde{L}_0$ decreases, the distribution becomes extremely super-thermal, reaching a limiting distribution of ${\rm d}\sigma / {\rm d} e_{\rm B} = e_{\rm B}(1+e_{\rm B})/\sqrt{1-e_{\rm B}^2}$ when $L_0 = 0$.  Our numerical scattering experiments reproduce this behavior qualitatively, with outcome distributions that are increasingly super-thermal as $L_0$ decreases.
\item The marginal distribution of binary orbital inclinations, ${\rm d}\sigma/{\rm d}C_{\rm B}$, likewise varies between two asymptotic limits.  When the ensemble $L_0=0$, there is no preferred direction, and this distribution is isotropic (${\rm d}\sigma/{\rm d}C_{\rm B} = 1/2$), as it must be by symmetry.  In most ensembles with realistic values of angular momentum, however, ${\rm d}\sigma/{\rm d}C_{\rm B}$ is strongly biased towards prograde orbits, which occupy more phase volume.  In this limit, ${\rm d}\sigma/{\rm d}C_{\rm B}$ is insensitive to both $\alpha$ and $L_0$.  The ergodic ${\rm d}\sigma/{\rm d}C_{\rm B}$ is well-matched by our numerical scattering experiments.
\end{itemize}
Overall agreement between numerical scattering experiments and the ergodic formalism we have developed here is good, indicating that the hypothesis of thermodynamic ergodicity is satisfied, at least for metastable triples that undergo multiple scrambles.  This conclusion, in combination with the large deviations from ergodic predictions seen in the $N_{\rm scram}=0$ subsamples, substantiates the intuitive supposition that it is {\it scrambles} that are responsible for generating chaotic orbital evolution and ``ergodicizing'' triple ensembles.  Even in the $N_{\rm scram} \ge 2$ subsamples we examine, however, we do see higher-order structure in marginal outcome distributions, beyond the leading-order agreement with our analytic predictions.  It is unclear whether this superimposed structure is generated by (i) islands of regularity in the space of initial conditions, or (ii) the crude idealizations that enter into our definition of the ``strong interaction region,'' and the need to employ a more realistic triple stability criterion.  In future work, we hope to better explore these possibilities.  We also hope to calibrate $\alpha$, the sole free parameter of our formalism, and to investigate the range of mass ratios $m_{\rm a}/m_{\rm s}$ and $m_{\rm a}/m_{\rm b}$ that yield ergodic behavior.  

One important caveat to this calculation is that we have examined the interactions of Newtonian point-particles.  In reality, finite-size effects (tidal forces or direct collisions) may play a role in the evolution of some {\it astrophysical} metastable triples, and if these triples contain compact objects, the dissipative nature of general relativistic gravity may also come to matter.  It would be interesting to extend our formalism to include these effects, though it may only be possible in an approximate manner.  We have also implicitly considered only hard binaries, in the sense that we assume there {\it is} a survivor binary.  If $E_0 > 0$, another possible outcome is a complete ionization, but in principle the techniques we have used here could be used to estimate distributions of (all three) escaper properties in such a scenario.

Generally, the dynamics and outcomes of binary-single encounters have been studied by numerical integration of the equations of motion, a process which can be time-consuming to do accurately.  Previous analytic work on three-body scatterings has generally made use of the impulse approximation to estimate cross-sections and outcomes for {\it promptly resolved} encounters, both in the soft binary\cite{Hut83b} and hard binary\cite{Heggie&Hut93} limits.  The closed-form outcome distributions predicted by our formalism provide an analytic understanding of {\it resonant} encounters, which in combination with past treatments of non-resonant scatterings, will allow theorists to replace large ensembles of numerical scattering experiments with reasonable accuracy.  This will aid in surveys of large parameter spaces for star cluster dynamics, and in the construction of analytic or semi-analytic models for the evolution of binary populations.  We hope it will also provide greater physical intuition into the evolution of these prototypical chaotic systems.

\end{methods}

%%%% ADDENDUM %%%%

\begin{addendum}
 \item[Data Availability Statement] The data that support the findings of this study are available from the corresponding author upon reasonable request.
\end{addendum}

% Extended data figures
\renewcommand{\figurename}{Extended Data Figure}
\setcounter{figure}{0}

\begin{figure*}
\centering
\includegraphics[width=155mm]{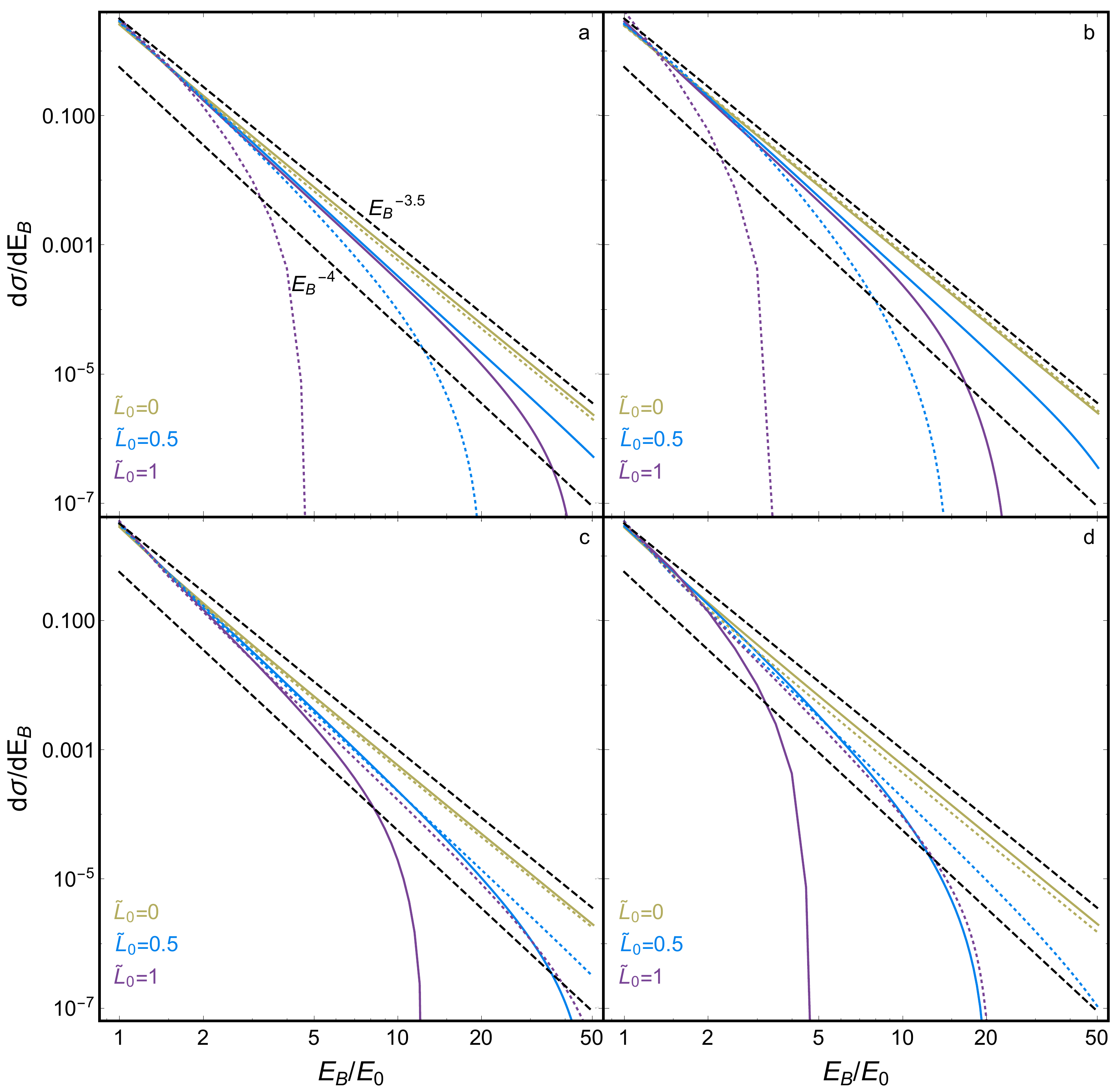}
\caption{{\bf The marginal distribution of binary energies, ${\rm d}\sigma/{\rm d}E_{\rm B}$}.  Colors show dimensionless angular momenta $\tilde{L}_0$ (but black dashed lines are analytic limits for $L_0 \ll 1$ and $L_0 \approx 1$).  {\bf a}: ergodic outcome distributions using the AE criterion, with $\alpha=2$, solid lines representing equal-mass scattering ensembles ($m_{\rm a}=m_{\rm b}=m_{\rm s}$), and dotted lines extreme mass-ratio ensembles ($m_{\rm a}=m_{\rm b}=10m_{\rm s}$).  {\bf b}: the same as top left, but for a SE criterion.  {\bf c}: intermediate-mass ratio scattering ensembles ($m_{\rm a}=m_{\rm b}=3m_{\rm s}$); solid lines are for $\alpha=2$, and dotted lines for $\alpha=5$.  {\bf d}: same as bottom left, but for $m_{\rm a}=m_{\rm b}=10m_{\rm s}$.}
\label{fig:marginalEnergy}
\end{figure*}

\begin{figure*}
\centering
\includegraphics[width=155mm]{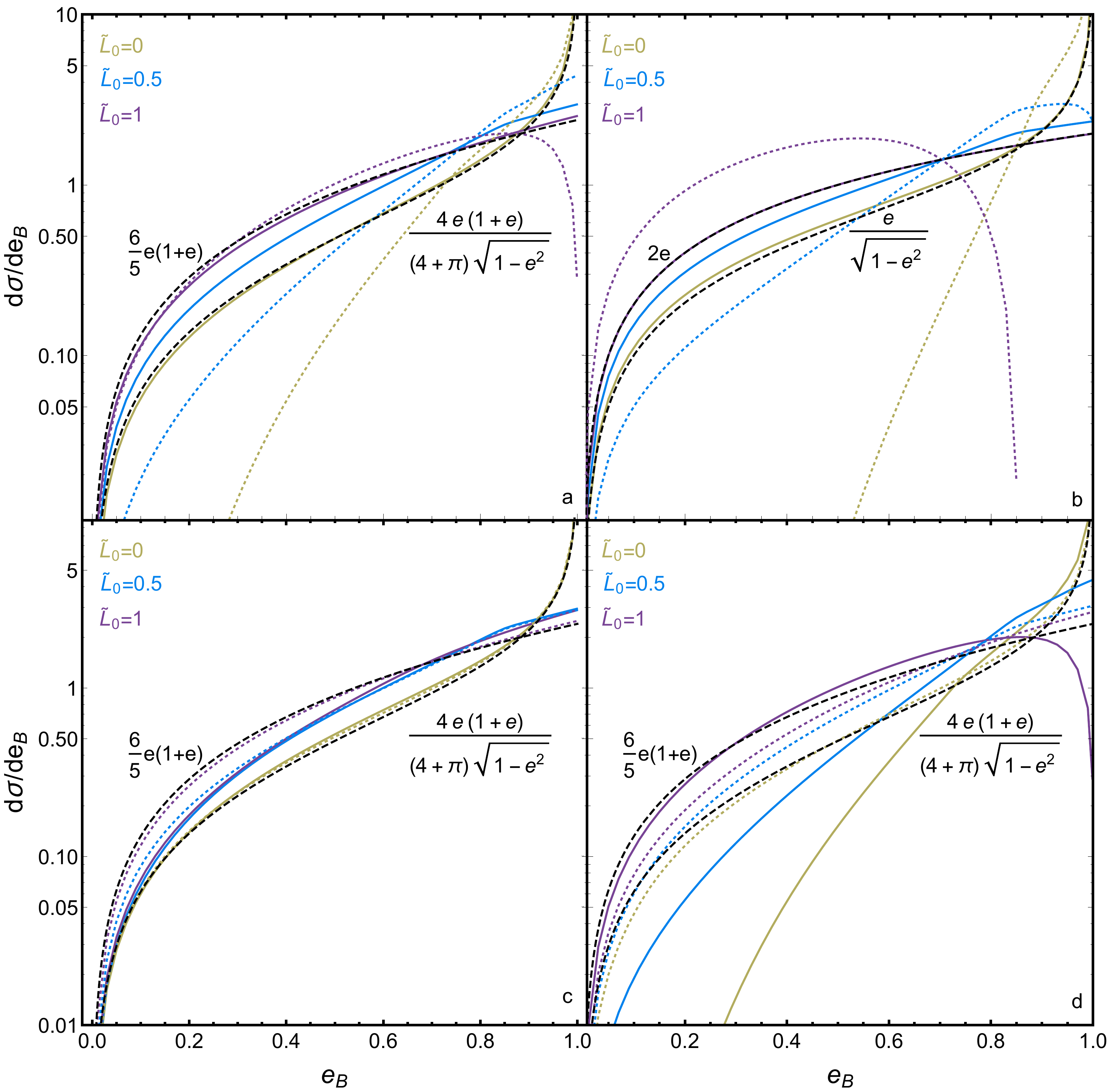}
\caption{{\bf The marginal distribution of binary eccentricity, ${\rm d}\sigma/{\rm d}e_{\rm B}$}.  Line styles, and assumptions in each panel are the same as in Extended Data Fig. \ref{fig:marginalEnergy}, except for black dashed lines, which here show $\tilde{L}_0\approx 1$ and $\tilde{L}_0 \ll 1$ limits of the ${\rm d}\sigma/{\rm d}e_{\rm B}$ distribution (unlike for ${\rm d}\sigma/{\rm d}E_{\rm B}$, these limits differ significantly in the AE and SE regimes).  In comparable-mass AE calculations, mildly super-thermal outcomes arise from geometric effects when $\tilde{L}_0 \sim 1$; in contrast, angular momentum starvation produces extremely super-thermal outcomes when $\tilde{L}_0 \ll 1$.  Small $m_{\rm s}$ values foreclose parts of $e_{\rm B}$ space, as $L_{\rm B} \approx L_0$.}
\label{fig:marginalEccentricity}
\end{figure*}

\begin{figure*}
\includegraphics[width=155mm]{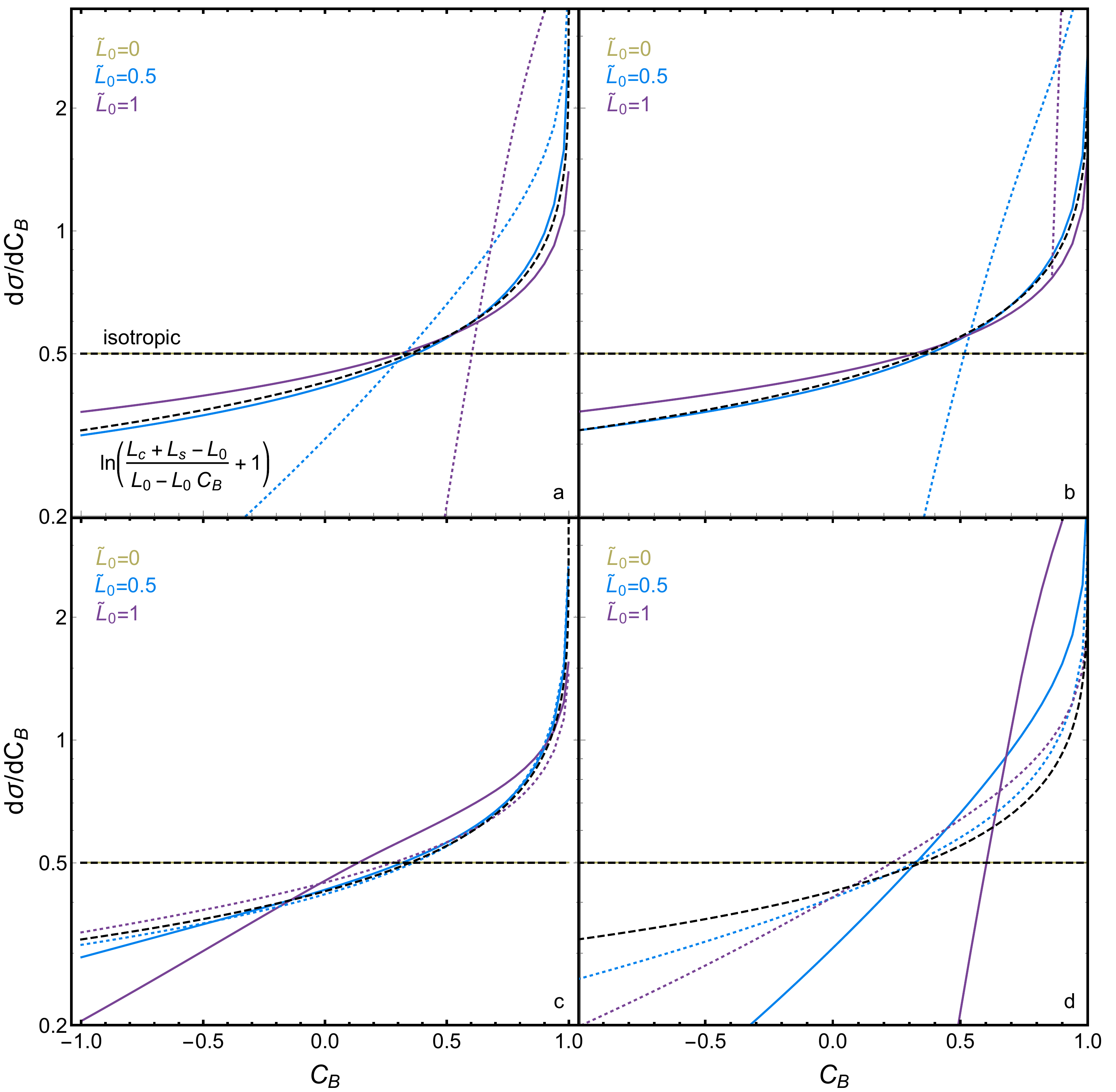}
\caption{{\bf The marginal distribution of binary orientation, ${\rm d}\sigma/{\rm d}C_{\rm B}$}.  Assumptions and line styles in each panel are the same as in Extended Data Fig. \ref{fig:marginalEnergy}, except that now the dashed black lines show (i) an isotropic outcome configuration, and (ii) an equal-mass, $\tilde{L}_0=0.5$ evaluation of Eq. \ref{eq:marginalC} (the $\tilde{L}_0 = 1.0$ evaluation is quite similar but is not shown for reasons of brevity).  For $\tilde{L}_0 \ll 1$, surviving binaries are distributed isotropically (as symmetry dictates).  Otherwise, binary orientations $C_{\rm B} = \hat{L}_{\rm B}\cdot \hat{L}_0$ are biased towards prograde outcomes.  For extreme mass ratios and large $\tilde{L}_0$, retrograde outcomes may be entirely prohibited.}
\label{fig:marginalInclination}
\end{figure*}

\clearpage
% Supplementary Table 1

\thispagestyle{empty}

\renewcommand{\tablename}{Extended Data Table}

\begin{table}
\centering
\begin{tabular}{ r || r | r | r | r | r}
  Run & $e_0$ & $\tilde{L}_0$ & $N_0$ & $N_1$ & $N_2$  \\
  \hline                        
  A & $0.0$ & 1.0 & 116,993 & 56,696 & 39,819  \\
  B & $0.5$ & 0.87 & 121,328 & 65,936 & 51,791  \\
  C & $0.9$ & 0.44 & 107,992 & 76,051 & 46,852  \\
          \end{tabular}
\caption{The numerical (binary-single) scattering ensembles we use for comparison to analytic theory in \S \ref{sec:numerics}.  The first two columns label the initial binary eccentricity $e_0$, and the conserved dimensionless angular momentum $\tilde{L}_0$ in each scattering run we simulate.  The latter columns show $N_{\rm i}$, the number of runs with $N_{\rm scram} \ge i$.  Each run has initial impact parameter $b=0$, isotropically distributed phase angles, and particles of equal mass ($m_{\rm a}=m_{\rm b}=m_{\rm s}$).}
\end{table}

\end{document}